\documentclass[aps,prb,twocolumn,showpacs,superscriptaddress]{revtex4-1}
\usepackage{amsmath}
\usepackage{amssymb}
\usepackage{graphicx}
\usepackage{color}
\usepackage{array}
\usepackage{makecell}

\begin{document}
\title{Possible Quantum Paraelectric State in Kitaev Spin Liquid Candidate H$_{3}$LiIr$_{2}$O$_{6}$}
\author{Shuai Wang}
\affiliation{International Center for Quantum Materials, School of Physics, Peking University, Beijing 100871, China}
\author{Long Zhang}
\email{longzhang@ucas.ac.cn}
\affiliation{Kavli Institute for Theoretical Sciences and CAS Center for Excellence in Topological Quantum Computation, University of Chinese Academy of Sciences, Beijing 100190, China}
\affiliation{Physical Science Laboratory, Huairou National Comprehensive Science Center, Beijing 101400, China}
\author{Fa Wang}
\email{wangfa@pku.edu.cn}
\affiliation{International Center for Quantum Materials, School of Physics, Peking University, Beijing 100871, China}
\affiliation{Collaborative Innovation Center of Quantum Matter, Beijing 100871, China}
\date{\today}

\begin{abstract}
A new quantum spin liquid (QSL) candidate material H$_{3}$LiIr$_{2}$O$_{6}$ was synthesized recently and was found not to show any magnetic order or phase transition down to low temperatures. In this work, we study the quantum dynamics of the hydrogen ions, i.e., protons, in this material by combining first-principles calculations and theoretical analysis. We show that each proton and its adjacent oxygen ions form an electric dipole. The dipole interactions and the proton tunneling are captured by a transverse-field Ising model with a quantum disordered paraelectric ground state. The dipole excitations have an energy gap $\Delta_{\mathrm{d}}\simeq 60$ meV, and can be probed by the infrared optical spectroscopy and the dielectric response. We argue that the electric dipole fluctuations renormalize the magnetic interactions in H$_{3}$LiIr$_{2}$O$_{6}$ and lead to a Kitaev QSL state.
\end{abstract}
\maketitle

\section{Introduction}

Quantum spin liquids (QSLs) are paramagnetic ground states of Mott insulators without any long-range magnetic orders or lattice symmetry breaking, which can be induced by (geometrical) frustration \cite{Balents2010} and strong charge fluctuations \cite{Zhou2017}. The QSLs are characterized by fractionalized spinons and emergent gauge flux excitations \cite{Wen2004quantum, Wen2002}. They were proposed to be the parent states of high-$T_{\mathrm{c}}$ superconductors \cite{Anderson1987, Lee2006} and may be used for quantum computation \cite{Kitaev2003, Kitaev2006a}.

\begin{figure}[tb]
\includegraphics[width=0.48\textwidth]{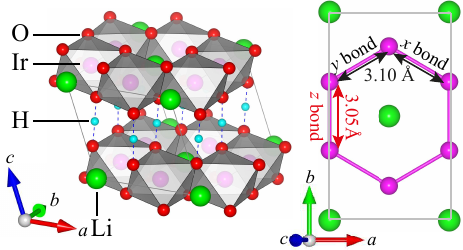}
\caption{(a) Crystal structure of H$_{3}$LiIr$_{2}$O$_{6}$ relaxed in first-principles calculations with the C2/m space group, i.e., the hydrogen atoms are restricted in the reflection plane by the space group. (b) The honeycomb lattice of iridium ions in the $ab$ plane. The three types of nearest-neighbor bonds in the Kitaev term are labeled.}
\label{fig:structure}
\end{figure}

The Kitaev model on the honeycomb lattice \cite{Kitaev2006a} is a prototype of QSL. The Hamiltonian hosts the bond-dependent Ising-type interactions,
\begin{equation}
H_{K}=K\sum_{\langle ij\rangle\in \gamma}S_{i}^{\gamma}S_{j}^{\gamma},
\end{equation}
where $\gamma=x, y, z$ labels the three types of nearest-neighbor bonds [Fig. \ref{fig:structure}, (b)]. The Kitaev model is exactly solvable with a QSL ground state \cite{Kitaev2006a}. Its excitations can be represented by Majorana fermions and emergent $\mathbb{Z}_{2}$ gauge fluxes.

It was soon realized \cite{Jackeli2009} that the Kitaev interaction naturally arises in several transition metal compounds, e.g., Na$_{2}$IrO$_{3}$, $\alpha$-Li$_{2}$IrO$_{3}$, and $\alpha$-RuCl$_{3}$ \cite{Plumb2014}. In these quasi-two dimensional materials, the edge-sharing IrO$_{6}$ (RuCl$_{6}$) octahedra form a honeycomb lattice in the $ab$ plane. The strong spin-orbit coupling on the cations lifts the degeneracy of the $t_{2g}$ orbitals and leaves a pseudospin $J_{\mathrm{eff}}=1/2$ Kramers pair occupied by one electron. The anion-mediated electron hopping projected in this subband is strongly suppressed due to the destructive interference of the two Ir-O-Ir hopping paths \cite{Jackeli2009}. The leading-order magnetic interaction involves the Hund coupling on the cations and has exactly the form of the Kitaev term \cite{Jackeli2009}.

However, all these Kitaev QSL candidates turn out to have long-range magnetic orders at low temperatures \cite{Liu2011, Choi2012, Ye2012, Sears2015, Johnson2015, Williams2016}. This can be accounted for by the nonnegligible Heisenberg interactions up to third nearest neighbors \cite{Kimchi2011},
\begin{equation} \label{eq:khmodel}
H_{J}=J_{1}\sum_{\langle ij\rangle}\vec{S}_{i}\cdot\vec{S}_{j}+J_{2}\sum_{\langle\langle ij\rangle\rangle}\vec{S}_{i}\cdot\vec{S}_{j} +J_{3}\sum_{\langle\langle\langle ij\rangle\rangle\rangle}\vec{S}_{i}\cdot\vec{S}_{j},
\end{equation}
and/or other spin-anisotropic interactions \cite{Rau2014a}.

A new Kitaev candidate material, H$_{3}$LiIr$_{2}$O$_{6}$, was synthesized recently by substituting hydrogen for the inter-IrO$_{3}$-layer lithium ions in $\alpha$-Li$_{2}$IrO$_{3}$ \cite{Kitagawa2018}. The nuclear magnetic resonance and thermodynamic measurements did not find any magnetic order or spin glassiness down to 50 mK despite a large Curie-Weiss temperature $\theta_{\mathrm{CW}}=-105$ K, thus suggesting a QSL state.

In this work, we study H$_{3}$LiIr$_{2}$O$_{6}$ with first-principles calculations and theoretical analysis, focusing on the role of the substitute hydrogen ions. We find that each hydrogen ion, i.e., proton, together with two adjacent oxygen ions, forms a uniaxial electric dipole almost perpendicular to the $ab$ plane. The electric dipole-dipole interaction is described by the Ising model on the ABC-stacking triangular lattice. The quantum tunneling of the proton flips the electric dipole and corresponds to a strong transverse field term in the Ising model, and leads to a quantum disordered paraelectric ground state. We thus predict a sizable dielectric response in a broad temperature range. The dipole excitations correspond to an optical phonon mode and can be probed by the infrared optical spectroscopy. We argue that the electric dipole fluctuations can renormalize the magnetic interaction parameters and may push the effective Kitaev-Heisenberg model of H$_{3}$LiIr$_{2}$O$_{6}$ into the Kitaev QSL phase.

\section{Crystal structure and electric dipoles on the O-H-O bonds}

Given the sensitivity of the magnetic properties of H$_{3}$LiIr$_{2}$O$_{6}$ to the hydrogen substitution, we first study the crystal structure with first-principles calculations and pay particular attention to the impact of the substitute hydrogen ions.

The first-principles calculations are performed with the projector augmented wave (PAW) method \cite{Blochl1994, Kresse1999} and the generalized gradient approximation (GGA) to the exchange-correlation functional \cite{Perdew1996}, which are implemented in the Vienna {\it ab initio} Simulation Package (VASP) \cite{Kresse1996, Kresse1996a}. The spin-orbit coupling and the onsite Coulomb interaction on the iridium atoms are included in the noncollinear magnetic calculations. The details of the implementation of first-principles calculations are presented in the Supplemental Materials.

We first adopt the experimental crystal structure of H$_{3}$LiIr$_{2}$O$_{6}$ without stacking faults refined with a monoclinic structure and the C2/m space group \cite{Bette2017}. The precise positions of the protons were not determined by the X-ray diffraction, therefore, their positions are fully relaxed and optimized in first-principles calculations until the force on each atom is smaller than 0.01 eV/\AA.

The relaxed crystal structure that preserves the C2/m symmetry is shown in Fig. \ref{fig:structure}. The hydrogen ions deviate from the original lithium positions in $\alpha$-Li$_{2}$IrO$_{3}$ and bridge the two nearest oxygen ions in the two adjacent IrO$_{3}$ layers, and form an ABC-stacking triangular lattice. The O-H-O bonds are almost perpendicular to the $ab$ plane.

\begin{figure}[tb]
\includegraphics[width=0.48\textwidth]{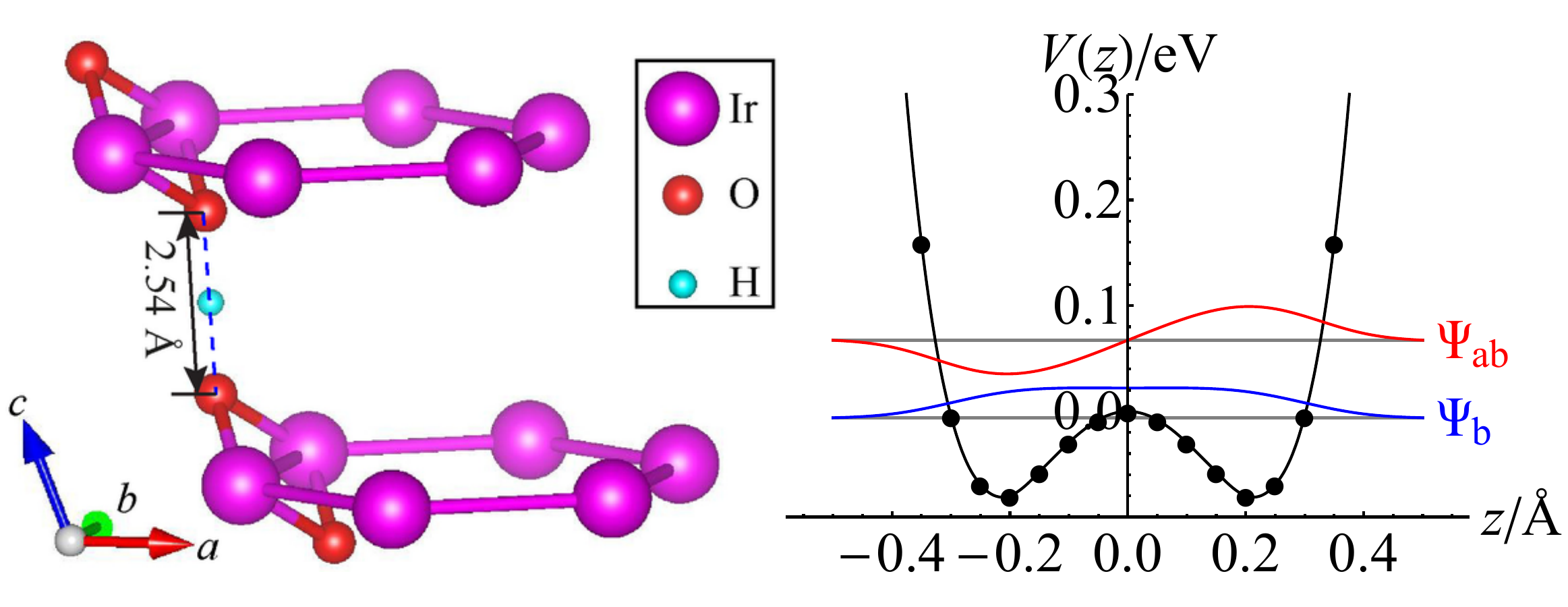}
\caption{(a) Schematic illustration of the O-H-O bond, which is almost perpendicular to the $ab$ plane. (b) The 1D double-well potential $V(z)$ along the O-H-O bond direction obtained by interpolating the first-principles energies with one proton placed at different heights (solid circles), where the origin is chosen at the bond center. The proton wavefunctions of the bonding ($\Psi_{\mathrm{b}}$) and the antibonding ($\Psi_{\mathrm{ab}}$) states are also shown.}
\label{fig:doublewell}
\end{figure}

The total energy is further lowered if the proton is shifted away from the O-H-O bond center towards either one of the oxygen ions, which breaks the C2/m symmetry, while its displacement in the $ab$ plane increases the energy significantly. By varying the height of one proton while fixing other protons at the O-H-O bond centers in a supercell in the first-principles calculations, we find that the proton is trapped in a double-well potential $V(z)$ [Fig. \ref{fig:doublewell}, (b)]. The potential minima are at $\pm 0.22$ {\AA} away from the O-H-O bond center. We note that there are two types of O-H-O bonds formed by O(2)-H(1)-O(2) and O(1)-H(2)-O(1) with the bond lengths $2.54$ {\AA} and $2.46$ {\AA}, respectively. The proton energy potentials and the electric dipole moments of them are different. The results presented in \ref{fig:doublewell} correspond to the O(2)-H(1)-O(2) bonds. The results of the O(1)-H(2)-O(1) bonds are presented in the Supplemental Materials.

The proton at one of the energy minima forms a uniaxial electric dipole with the oxygen ions. The net dipole moment is calculated by integrating the dipole moment density over a cylinder surrounding the O-H-O bond, $p_{0}=0.06\sim 0.11~e\cdot\mathrm{\AA}$. (The uncertainty comes from different choices of the cylinder height.)

\section{Electric dipole-dipole interactions}

\begin{figure}[tb]
\includegraphics[width=0.48\textwidth]{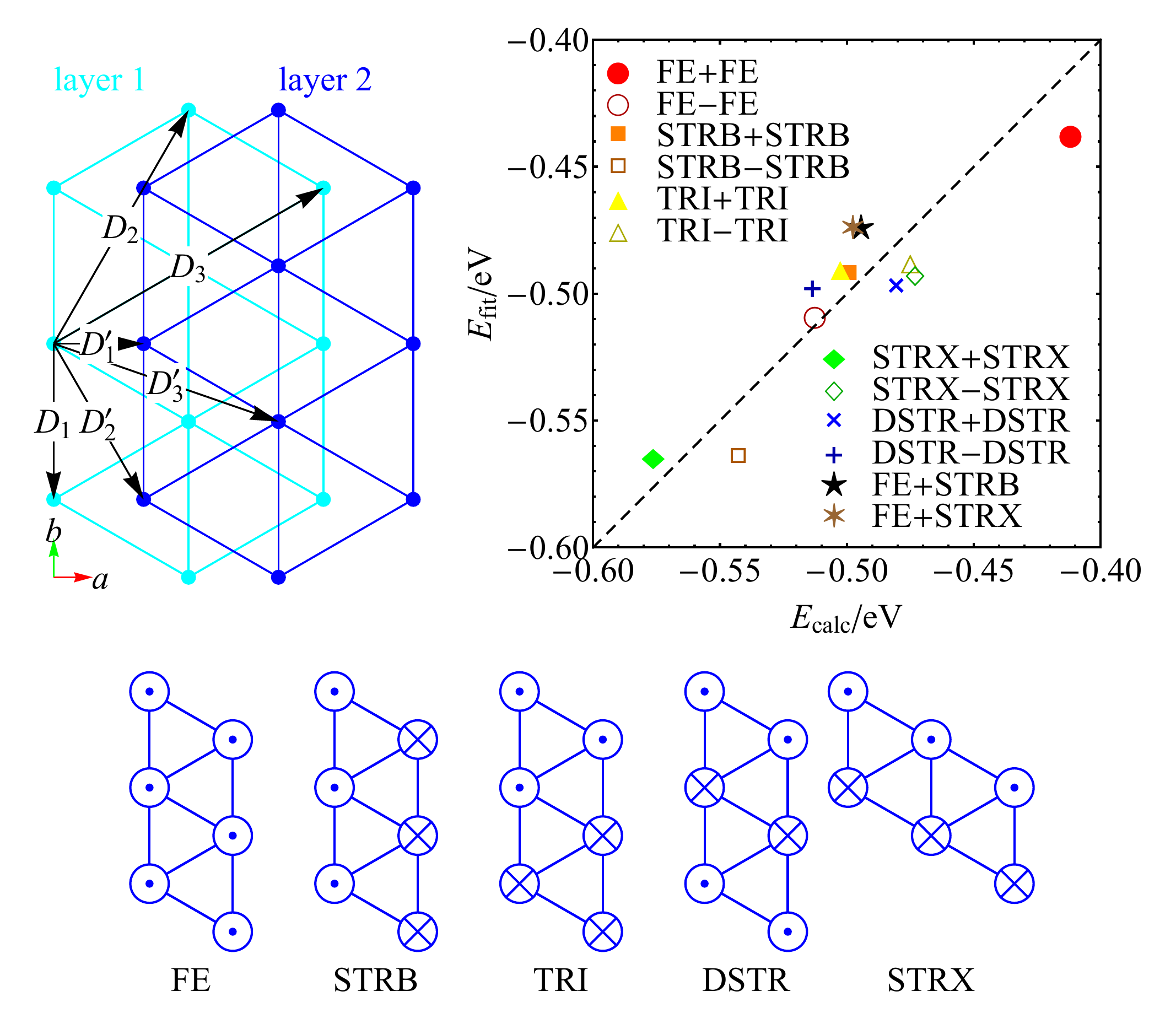}
\caption{(a) ABC-stacking triangular lattice of the electric dipoles. The Ising interaction parameters $D_{1,2,3}$ (intralayer) and $D'_{1,2,3}$ (interlayer) are labeled on the corresponding bonds. (b) The comparison of the total energies from first-principles calculations $E_{\mathrm{calc}}$ (horizontal axis) and the interaction energies in the Ising model $E_{\mathrm{fit}}$ (vertical axis). The Ising interaction parameters are obtained by the least-square fitting. (c) Several intralayer dipole configurations used in calculations. The label ``FE$+$FE'' (``FE$-$FE'') indicates that one of the two layers has the same dipole configuration (has all dipole directions reversed) as the other layer. The dots and the crosses in the schematic dipole configurations stand for the electric dipoles pointing upward or downward, respectively.}
\label{fig:ising}
\end{figure}

We treat these uniaxial electric dipoles as Ising variables, $\sigma_{i}^{z}=\pm 1$. In order to capture the dipole-dipole interactions in the crystal, we obtain the total energies of various dipole configurations from the first-principles calculations (Fig. \ref{fig:ising}) and fit the Ising model on the ABC-stacking triangular lattice,
\begin{equation}
H_{D}=\sum_{ij}D_{ij}\sigma_{i}^{z}\sigma_{j}^{z},
\end{equation}
where $D_{ij}$ denotes both the intralayer interactions $D_{1,2,3}$ and the interlayer interactions $D'_{1,2,3}$ up to the third nearest neighbors defined in Fig. \ref{fig:ising}. This method was adopted to study the electric dipole-dipole interactions in the hexaferrite BaFe$_{12}$O$_{19}$ \cite{Wang2014f}. The least-square fitting yields $D_{1}=1.7$ meV, $D_{2}=-0.5$ meV, $D_{3}=-0.3$ meV, $D_{1}'=-0.2$ meV, $D_{2}'=1.0$ meV, and $D_{3}'=0.1$ meV. The details are presented in the Supplemental Materials. The comparison of the first-principles total energies and the fitted model energies is shown in Fig. \ref{fig:ising}. On the other hand, the intralayer nearest-neighbor interaction can be estimated from the dipole-dipole interaction at a distance $r=3.15$ \AA, $D_{1}\simeq p_{0}^{2}/4\pi\varepsilon_{0}r^{3}=1.7\sim 5.6$ meV, which is roughly consistent with the result of fitting.

\section{Proton tunneling and quantum paraelectricity}

The Ising model of the dipole-dipole interactions may suggest an antiferroelectric order at the ground state; however, we will show this is not the case because of the quantum tunneling of the protons. The proton tunneling between the energy minima flips the electric dipole, and thus acts as a transverse field in the Ising model,
\begin{equation}
\label{eq:hx}
H_{h}=-h_{x}\sum_{i}\sigma_{i}^{x}.
\end{equation}
$h_{x}$ can be calculated from the difference of the bonding and the antibonding state energies, $\epsilon_{\mathrm{b}}$ and $\epsilon_{\mathrm{ab}}$, in the double-well potential $V(z)$. By numerically solving the 1D Schr\"odinger equation,
\begin{equation}
-\frac{\hbar^{2}}{2m_{\mathrm{p}}}\psi''(z)+V(z)\psi(z)=\epsilon \psi(z),
\end{equation}
where $m_{\mathrm{p}}$ is the proton mass, we find $h_{x}=(\epsilon_{\mathrm{ab}}-\epsilon_{\mathrm{b}})/2=36.7$ meV. The wavefunctions of the bonding and the antibonding states are shown in Fig. \ref{fig:doublewell}.

The proton tunneling term dominates over the dipole interactions, $h_{x}\gg |D_{ij}|$, therefore the ground state of the electric dipoles is a quantum disordered paraelectric state and the C2/m symmetry is thus dynamically restored. This leads to the following predictions to experiments. First, these electric dipoles contribute a sizable uniaxial dielectric response to the electric field perpendicular to the $ab$ plane in a large temperature range. The temperature-dependence of the electric susceptibility $\chi_{e}(T)$ can be derived by the mean field theory of the transverse-field Ising model, and the result is the Barrett formula \cite{Barrett1952},
\begin{equation}
\chi_{e}(T)=\frac{M}{\frac{1}{2}T_{1}\coth (T_{1}/2T)-T_{0}},
\end{equation}
where $T_{1}=2h_{x}/k_{\mathrm{B}}\simeq 870$ K marks the crossover from the high-$T$ Curie-Weiss behavior to the low-$T$ plateau, $\chi_{e}(T\rightarrow 0)=M/(T_{1}/2-T_{0})$, and $T_{0}=-k_{\mathrm{B}}^{-1}\sum_{j}D_{ij}\simeq -130$ K is the effective antiferroelectric interaction strength. The overall amplitude $M=\rho_{0}p_{0}^{2}/\varepsilon_{0}k_{\mathrm{B}}$ depends on the electric dipole moment $p_{0}$ and the dipole density $\rho_{0}$.

\begin{figure}[tb]
\includegraphics[width=0.48\textwidth]{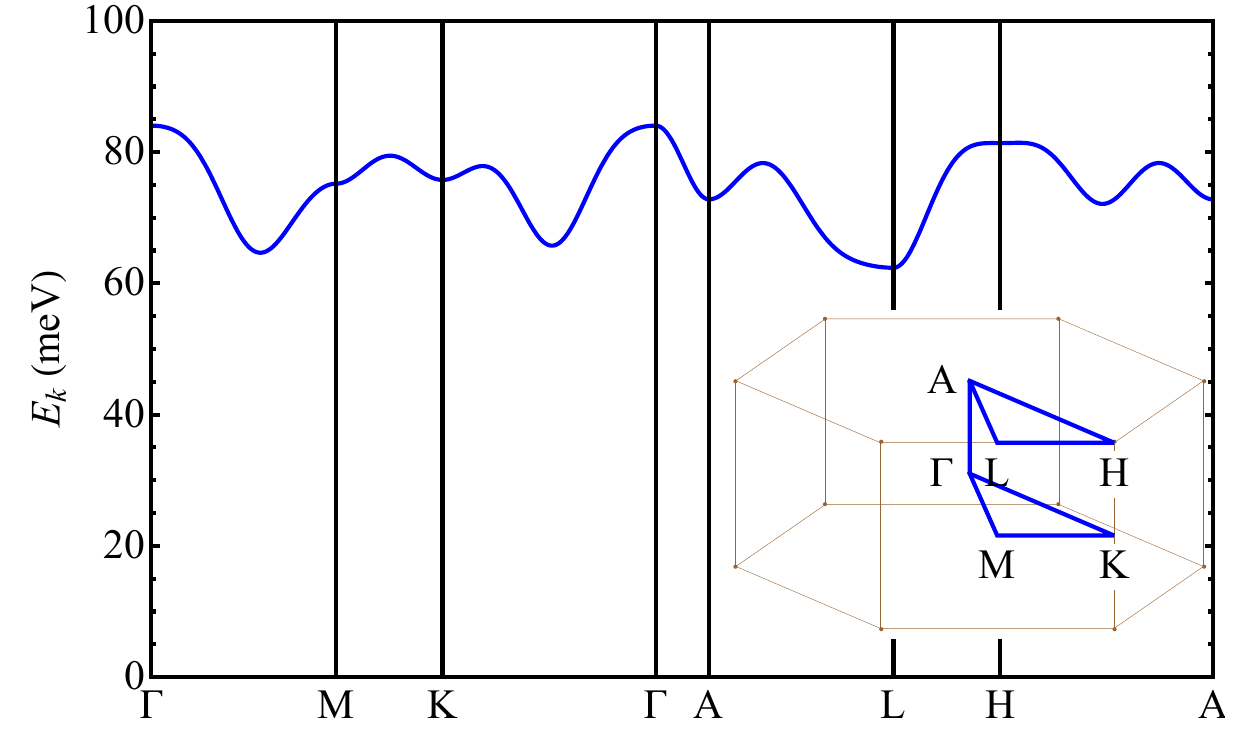}
\caption{The dipole excitation spectrum along the high-symmetry lines in the momentum space (inset).}
\label{fig:excitation}
\end{figure}

Second, the electric dipole excitations correspond to an optical phonon mode. Its spectrum is derived by a single-mode approximation on the paraelectric ground state and is shown in Fig. \ref{fig:excitation}. There is a dipole excitation gap $\Delta_{\mathrm{d}}\simeq 60$ meV. These dipole excitations may be probed by the infrared optical spectroscopy.

\section{Discussion: Impact on the magnetic interactions}

The magnetic interactions can be extracted from the total energies of various magnetic moment configurations when both the atom positions and the magnetic moment configurations are prescribed in the first-principles calculations. The details are presented in the Supplemental Materials. The protons are placed at the O-H-O bond centers preserving the C2/m symmetry. The following (extended) Kitaev-Heisenberg model is fitted to the calculated total energies of various magnetic configurations,
\begin{equation}
\begin{split}
H_{\mathrm{KH}}&= H_{K}+H_{J} +\Gamma\sum_{\langle ij\rangle\in\gamma}(S_{i}^{\alpha}S_{j}^{\beta}+S_{i}^{\beta}S_{j}^{\alpha})\\
&+\Gamma'\sum_{\langle ij\rangle\in\gamma}(S_{i}^{\alpha}S_{j}^{\gamma}+S_{i}^{\gamma}S_{j}^{\alpha}+S_{i}^{\beta}S_{j}^{\gamma}+S_{i}^{\gamma}S_{j}^{\beta}),
\end{split}
\end{equation}
where $(\alpha,\beta,\gamma)$ is the cyclic permutation of $(x, y, z)$. The fitted parameters are listed in Table \ref{tab:parameters}. The Curie-Weiss temperature of this model is $\theta_{\mathrm{CW}}=-K-3J_{1}-6J_{2}-3J_{3}=-104$ K. The results of the closely related materials Na$_{2}$IrO$_{3}$ and $\alpha$-Li$_{2}$IrO$_{3}$ derived from first-principles calculations \cite{Hu2015b} and nonperturbative exact diagonalization \cite{Winter2016} are also listed in Table \ref{tab:parameters} for comparison.

\begin{ruledtabular}
\begin{table}[b]
\caption{The magnetic interaction parameters (in meV) of H$_{3}$LiIr$_{2}$O$_{6}$ derived from first-principles calculations with the hydrogen ions placed at the O-H-O bond centers. The results of Na$_{2}$IrO$_{3}$ and $\alpha$-Li$_{2}$IrO$_{3}$ reported in Refs. \cite{Hu2015b} and \cite{Winter2016} are included for comparison. The parameters cited from Ref. \cite{Winter2016} are averaged over the three bond directions.}
\label{tab:parameters}
\begin{tabular}{c|cccccc}
Material 									& $K$ 	& $J_{1}$ 	& $J_{2}$ 	& $J_{3}$ 	& $\Gamma$ 	& $\Gamma'$ \\
\hline
H$_{3}$LiIr$_{2}$O$_{6}$ 					& $-21.6$	& $6.3$ 	& $0.4$ 	& $3.1$		& $-0.2$	& $-4.1$	\\
Na$_{2}$IrO$_{3}$ \cite{Hu2015b}			& $-19.1$	& $7.2$		& $-1.6$	& $7.8$		& $1.5$		& $-3.5$	\\
Na$_{2}$IrO$_{3}$ \cite{Winter2016} 		& $-16.8$	& $0.5$		& $0.2$		& $6.7$		& $1.4$		& $-2.1$	\\
$\alpha$-Li$_{2}$IrO$_{3}$ \cite{Winter2016}& $-8.6$	& $-2.7$	& $0.4$		& $6.0$		& $8.9$		& $-0.6$	\\
\end{tabular}
\end{table}
\end{ruledtabular}

\begin{figure}[tb]
\includegraphics[width=0.48\textwidth]{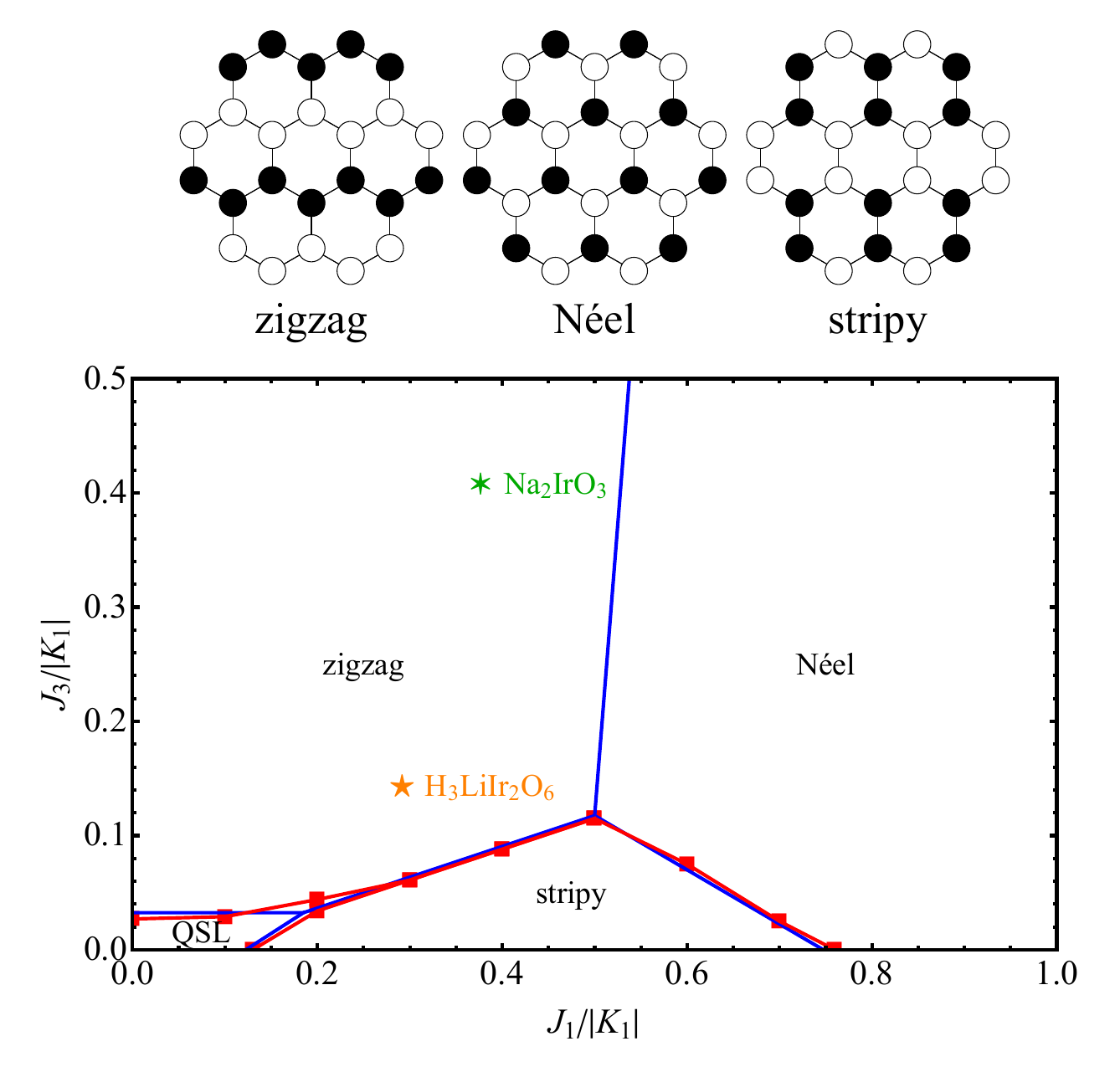}
\caption{Phase diagram of the $K$-$J_{1}$-$J_{3}$ model obtained by exact diagonalization on a 24-site cluster (blue lines) and on a 32-site cluster (red lines). The parameters of Na$_{2}$IrO$_{3}$ and H$_{3}$LiIr$_{2}$O$_{6}$ are labeled.}
\label{fig:phasediag}
\end{figure}

It is instructive to compare the magnetic interaction parameters of H$_{3}$LiIr$_{2}$O$_{6}$ and Na$_{2}$IrO$_{3}$. In both materials, the spin-anisotropic $\Gamma$ and $\Gamma'$ terms as well as the $J_{2}$ term are relatively small, which suggests a minimal model incorporating only the $K$-$J_{1}$-$J_{3}$ terms \cite{Winter2016}, which is not applicable to $\alpha$-Li$_{2}$IrO$_{3}$ due to a large $\Gamma$ term. The phase diagram of this model is obtained by exact diagonalization on small lattice clusters (Fig. \ref{fig:phasediag}) \cite{Winter2016}. Both $J_{1}$ and $J_{3}$ terms can destabilize the Kitaev QSL phase and lead to magnetic ordered states. In particular, a sizable $J_{3}$ term favors the ziazag order, which is consistent with the experiments on Na$_{2}$IrO$_{3}$ \cite{Liu2011, Choi2012, Ye2012}. Both ratios $J_{1}/|K|$ and $J_{3}/|K|$ in H$_{3}$LiIr$_{2}$O$_{6}$ are significantly reduced upon the hydrogen substitution, thus its ground state is closer to the Kitaev QSL phase. Nevertheless, the fitted $K$-$J_1$-$J_3$ model of H$_{3}$LiIr$_{2}$O$_{6}$ remains in the zigzag ordered phase.

We argue that the proton fluctuations may further push the effective magnetic interaction parameters of H$_{3}$LiIr$_{2}$O$_{6}$ into the Kitaev QSL phase. Recall that the oxygen-mediated nearest-neighbor electron hopping in the Slater-Koster approximation exactly vanishes in the pseudospin $J_{\mathrm{eff}}=1/2$ subspace due to the destructive interference of the two Ir-O-Ir hopping paths \cite{Jackeli2009}. This exact cancellation is absent when the fluctuations of the nearby protons set in and the two hopping paths are not equivalent. When the proton gets closer to one of the oxygen ions, the onsite energy of a hole at this oxygen ion is increased, $E_{p}\rightarrow E_{p}+\delta E_{p}$, thus the hopping amplitude along this path is reduced, $t\rightarrow t-\delta t$, with $\delta t/t\simeq \delta E_{p}/E_{p}$, and the two hopping paths do not cancel out completely. For example, on a $z$-bond, the remnant hopping term is
\begin{equation}
\delta H_{t}=-\frac{1}{3}\delta t (\sigma_{1}^{z}+\sigma_{2}^{z}) \sum_{\alpha}i\alpha d_{j\alpha}^{\dagger}d_{i\alpha}+\mathrm{h.c.},
\end{equation}
where $d_{j\alpha}^{\dagger}$ is the creation operator of the pseudospin $J_{\mathrm{eff}}=1/2$ electron on the iridium ion, and $\sigma_{1}^{z}$ and $\sigma_{2}^{z}$ are the configurations of the two dipoles close to the two oxygen ions, respectively. Treating it as a perturbation gives the following correction to the effective magnetic interactions,
\begin{equation} \label{eq:deltah}
\delta H_{\mathrm{KH}}=\delta J\sum_{\langle ij\rangle\in \gamma}(2S_{i}^{\gamma}S_{j}^{\gamma}-\vec{S}_{i}\cdot \vec{S}_{j}),
\end{equation}
where $\delta J =8\delta t^{2}/9(U+2h_{x})$ and $U$ is the onsite repulsion on the iridium ions. Therefore, the proton fluctuations renormalize the magnetic interaction parameters, $|K|\rightarrow |K|-2\delta J$, and $J_{1}\rightarrow J_{1}-\delta J$. The ratio $J_{1}/|K|$ is nevertheless reduced because $J_{1}/|K|<0.5$. By considering the oxygen-mediated long-range hopping paths, we may similarly argue that $J_{3}$ is also reduced by the proton fluctuations. Therefore, the proton/electric dipole fluctuations may further push the magnetic interactions towards the Kitaev QSL phase.

\section{Summary and outlook}

We have studied the physical consequences of the hydrogen substitution in the Kitaev QSL candidate H$_{3}$LiIr$_{2}$O$_{6}$. We find that each proton is trapped in a double-well potential and forms an electric dipole with two adjacent oxygen ions. Incorporating the dipole interactions and the proton tunneling, the low-energy dynamics of these dipoles is captured by a transverse-field Ising model. The strong proton tunneling leads to a quantum disordered paraelectric ground state. The dipole excitations may be probed with the dielectric response and the infrared optical spectroscopy. The dipole fluctuations renormalize the effective magnetic interactions and may push the magnetic ground state of H$_{3}$LiIr$_{2}$O$_{6}$ into the Kitaev QSL phase. Similar study of the role of the hydrogen ions in H$_{3}$LiIr$_{2}$O$_{6}$ was also presented in two independent works published recently \cite{Li2018a, Yadav2018}. Evidence of the proposed quantum paraelectricity has been obtained in recent experiments with dielectric spectroscopy \cite{Geirhos2020}.

The electric dipoles formed by displaced ions are ubiquitous in materials. Our approach to the quantum dynamics of the electric dipoles can be applied to other materials. Moreover, the interplay of the electric dipoles and the spin and orbit degrees of freedom may provide a new route towards quantum paraelectric states, QSLs and other exotic quantum states of matter \cite{Shen2016a, Shimozawa2017, Hassan2018, Zheng2018}.

In this work, we have not analyzed the unusual thermodynamic behaviors of H$_{3}$LiIr$_{2}$O$_{6}$ found in the experiments \cite{Kitagawa2018}.
In a recent theoretical work accounting for this behavior \cite{Slagle2018} was based on the assumption that a single layer of H$_{3}$LiIr$_{2}$O$_{6}$ would be in the Kitaev QSL phase. On the other hand, it was also proposed that the low-energy excitations could be attributed to a random singlet phase \cite{Kimchi2018}, which is qualitative consistent with the recent experimental evidence of slowing-down dynamics in the dielectric response at low temperature \cite{Geirhos2020}. The effect of quenched disorder on the thermodynamic and magnetic properties in H$_{3}$LiIr$_{2}$O$_{6}$ and other related materials is worth further study.

\acknowledgements
We are grateful to inspiring discussions with Xingye Lu and Weiqiang Yu. The numerical simulations were performed on Tianhe-I Supercomputer Systerm in Tianjin and on Tianhe-II Supercomputer System in Guangzhou. This work is supported by National Key Basic Research Program of China (No. 2014CB920902), National Key R\&D Program of China (Nos. 2017YFA0302904 and 2018YFA0305800), National Natural Science Foundation of China (No. 11804337), Strategic Priority Research Program of Chinese Academy of Sciences (No. XDB28000000), and Beijing Municipal Science \& Technology Commission (No. Z181100004218001).

\bibliography{library,books}

\begin{thebibliography}{38}%
\makeatletter
\providecommand \@ifxundefined [1]{%
 \@ifx{#1\undefined}
}%
\providecommand \@ifnum [1]{%
 \ifnum #1\expandafter \@firstoftwo
 \else \expandafter \@secondoftwo
 \fi
}%
\providecommand \@ifx [1]{%
 \ifx #1\expandafter \@firstoftwo
 \else \expandafter \@secondoftwo
 \fi
}%
\providecommand \natexlab [1]{#1}%
\providecommand \enquote  [1]{``#1''}%
\providecommand \bibnamefont  [1]{#1}%
\providecommand \bibfnamefont [1]{#1}%
\providecommand \citenamefont [1]{#1}%
\providecommand \href@noop [0]{\@secondoftwo}%
\providecommand \href [0]{\begingroup \@sanitize@url \@href}%
\providecommand \@href[1]{\@@startlink{#1}\@@href}%
\providecommand \@@href[1]{\endgroup#1\@@endlink}%
\providecommand \@sanitize@url [0]{\catcode `\\12\catcode `\$12\catcode
  `\&12\catcode `\#12\catcode `\^12\catcode `\_12\catcode `\%12\relax}%
\providecommand \@@startlink[1]{}%
\providecommand \@@endlink[0]{}%
\providecommand \url  [0]{\begingroup\@sanitize@url \@url }%
\providecommand \@url [1]{\endgroup\@href {#1}{\urlprefix }}%
\providecommand \urlprefix  [0]{URL }%
\providecommand \Eprint [0]{\href }%
\providecommand \doibase [0]{http://dx.doi.org/}%
\providecommand \selectlanguage [0]{\@gobble}%
\providecommand \bibinfo  [0]{\@secondoftwo}%
\providecommand \bibfield  [0]{\@secondoftwo}%
\providecommand \translation [1]{[#1]}%
\providecommand \BibitemOpen [0]{}%
\providecommand \bibitemStop [0]{}%
\providecommand \bibitemNoStop [0]{.\EOS\space}%
\providecommand \EOS [0]{\spacefactor3000\relax}%
\providecommand \BibitemShut  [1]{\csname bibitem#1\endcsname}%
\let\auto@bib@innerbib\@empty
\bibitem [{\citenamefont {Balents}(2010)}]{Balents2010}%
  \BibitemOpen
  \bibfield  {author} {\bibinfo {author} {\bibfnamefont {L.}~\bibnamefont
  {Balents}},\ }\href {\doibase 10.1038/nature08917} {\bibfield  {journal}
  {\bibinfo  {journal} {Nature}\ }\textbf {\bibinfo {volume} {464}},\ \bibinfo
  {pages} {199} (\bibinfo {year} {2010})}\BibitemShut {NoStop}%
\bibitem [{\citenamefont {Zhou}\ \emph {et~al.}(2017)\citenamefont {Zhou},
  \citenamefont {Kanoda},\ and\ \citenamefont {Ng}}]{Zhou2017}%
  \BibitemOpen
  \bibfield  {author} {\bibinfo {author} {\bibfnamefont {Y.}~\bibnamefont
  {Zhou}}, \bibinfo {author} {\bibfnamefont {K.}~\bibnamefont {Kanoda}}, \ and\
  \bibinfo {author} {\bibfnamefont {T.-K.}\ \bibnamefont {Ng}},\ }\href
  {\doibase 10.1103/RevModPhys.89.025003} {\bibfield  {journal} {\bibinfo
  {journal} {Rev. Mod. Phys.}\ }\textbf {\bibinfo {volume} {89}},\ \bibinfo
  {pages} {025003} (\bibinfo {year} {2017})}\BibitemShut {NoStop}%
\bibitem [{\citenamefont {Wen}(2004)}]{Wen2004quantum}%
  \BibitemOpen
  \bibfield  {author} {\bibinfo {author} {\bibfnamefont {X.-G.}\ \bibnamefont
  {Wen}},\ }\href@noop {} {\emph {\bibinfo {title} {{Quantum field theory of
  many-body systems: from the origin of sound to an origin of light and
  electrons}}}}\ (\bibinfo  {publisher} {Oxford University Press},\ \bibinfo
  {year} {2004})\BibitemShut {NoStop}%
\bibitem [{\citenamefont {Wen}(2002)}]{Wen2002}%
  \BibitemOpen
  \bibfield  {author} {\bibinfo {author} {\bibfnamefont {X.-G.}\ \bibnamefont
  {Wen}},\ }\href {\doibase 10.1103/PhysRevB.65.165113} {\bibfield  {journal}
  {\bibinfo  {journal} {Phys. Rev. B}\ }\textbf {\bibinfo {volume} {65}},\
  \bibinfo {pages} {165113} (\bibinfo {year} {2002})}\BibitemShut {NoStop}%
\bibitem [{\citenamefont {Anderson}(1987)}]{Anderson1987}%
  \BibitemOpen
  \bibfield  {author} {\bibinfo {author} {\bibfnamefont {P.~W.}\ \bibnamefont
  {Anderson}},\ }\href {\doibase 10.1126/science.235.4793.1196} {\bibfield
  {journal} {\bibinfo  {journal} {Science}\ }\textbf {\bibinfo {volume}
  {235}},\ \bibinfo {pages} {1196} (\bibinfo {year} {1987})}\BibitemShut
  {NoStop}%
\bibitem [{\citenamefont {Lee}\ \emph {et~al.}(2006)\citenamefont {Lee},
  \citenamefont {Nagaosa},\ and\ \citenamefont {Wen}}]{Lee2006}%
  \BibitemOpen
  \bibfield  {author} {\bibinfo {author} {\bibfnamefont {P.~A.}\ \bibnamefont
  {Lee}}, \bibinfo {author} {\bibfnamefont {N.}~\bibnamefont {Nagaosa}}, \ and\
  \bibinfo {author} {\bibfnamefont {X.-G.}\ \bibnamefont {Wen}},\ }\href
  {\doibase 10.1103/RevModPhys.78.17} {\bibfield  {journal} {\bibinfo
  {journal} {Rev. Mod. Phys.}\ }\textbf {\bibinfo {volume} {78}},\ \bibinfo
  {pages} {17} (\bibinfo {year} {2006})}\BibitemShut {NoStop}%
\bibitem [{\citenamefont {Kitaev}(2003)}]{Kitaev2003}%
  \BibitemOpen
  \bibfield  {author} {\bibinfo {author} {\bibfnamefont {A.}~\bibnamefont
  {Kitaev}},\ }\href {\doibase 10.1016/S0003-4916(02)00018-0} {\bibfield
  {journal} {\bibinfo  {journal} {Ann. Phys. (N. Y).}\ }\textbf {\bibinfo
  {volume} {303}},\ \bibinfo {pages} {2} (\bibinfo {year} {2003})}\BibitemShut
  {NoStop}%
\bibitem [{\citenamefont {Kitaev}(2006)}]{Kitaev2006a}%
  \BibitemOpen
  \bibfield  {author} {\bibinfo {author} {\bibfnamefont {A.}~\bibnamefont
  {Kitaev}},\ }\href {\doibase 10.1016/j.aop.2005.10.005} {\bibfield  {journal}
  {\bibinfo  {journal} {Ann. Phys. (N. Y).}\ }\textbf {\bibinfo {volume}
  {321}},\ \bibinfo {pages} {2} (\bibinfo {year} {2006})}\BibitemShut {NoStop}%
\bibitem [{\citenamefont {Jackeli}\ and\ \citenamefont
  {Khaliullin}(2009)}]{Jackeli2009}%
  \BibitemOpen
  \bibfield  {author} {\bibinfo {author} {\bibfnamefont {G.}~\bibnamefont
  {Jackeli}}\ and\ \bibinfo {author} {\bibfnamefont {G.}~\bibnamefont
  {Khaliullin}},\ }\href {\doibase 10.1103/PhysRevLett.102.017205} {\bibfield
  {journal} {\bibinfo  {journal} {Phys. Rev. Lett.}\ }\textbf {\bibinfo
  {volume} {102}},\ \bibinfo {pages} {017205} (\bibinfo {year}
  {2009})}\BibitemShut {NoStop}%
\bibitem [{\citenamefont {Plumb}\ \emph {et~al.}(2014)\citenamefont {Plumb},
  \citenamefont {Clancy}, \citenamefont {Sandilands}, \citenamefont {Shankar},
  \citenamefont {Hu}, \citenamefont {Burch}, \citenamefont {Kee},\ and\
  \citenamefont {Kim}}]{Plumb2014}%
  \BibitemOpen
  \bibfield  {author} {\bibinfo {author} {\bibfnamefont {K.~W.}\ \bibnamefont
  {Plumb}}, \bibinfo {author} {\bibfnamefont {J.~P.}\ \bibnamefont {Clancy}},
  \bibinfo {author} {\bibfnamefont {L.~J.}\ \bibnamefont {Sandilands}},
  \bibinfo {author} {\bibfnamefont {V.~V.}\ \bibnamefont {Shankar}}, \bibinfo
  {author} {\bibfnamefont {Y.~F.}\ \bibnamefont {Hu}}, \bibinfo {author}
  {\bibfnamefont {K.~S.}\ \bibnamefont {Burch}}, \bibinfo {author}
  {\bibfnamefont {H.-Y.}\ \bibnamefont {Kee}}, \ and\ \bibinfo {author}
  {\bibfnamefont {Y.-J.}\ \bibnamefont {Kim}},\ }\href {\doibase
  10.1103/PhysRevB.90.041112} {\bibfield  {journal} {\bibinfo  {journal} {Phys.
  Rev. B}\ }\textbf {\bibinfo {volume} {90}},\ \bibinfo {pages} {041112}
  (\bibinfo {year} {2014})}\BibitemShut {NoStop}%
\bibitem [{\citenamefont {Liu}\ \emph {et~al.}(2011)\citenamefont {Liu},
  \citenamefont {Berlijn}, \citenamefont {Yin}, \citenamefont {Ku},
  \citenamefont {Tsvelik}, \citenamefont {Kim}, \citenamefont {Gretarsson},
  \citenamefont {Singh}, \citenamefont {Gegenwart},\ and\ \citenamefont
  {Hill}}]{Liu2011}%
  \BibitemOpen
  \bibfield  {author} {\bibinfo {author} {\bibfnamefont {X.}~\bibnamefont
  {Liu}}, \bibinfo {author} {\bibfnamefont {T.}~\bibnamefont {Berlijn}},
  \bibinfo {author} {\bibfnamefont {W.-G.}\ \bibnamefont {Yin}}, \bibinfo
  {author} {\bibfnamefont {W.}~\bibnamefont {Ku}}, \bibinfo {author}
  {\bibfnamefont {A.}~\bibnamefont {Tsvelik}}, \bibinfo {author} {\bibfnamefont
  {Y.-J.}\ \bibnamefont {Kim}}, \bibinfo {author} {\bibfnamefont
  {H.}~\bibnamefont {Gretarsson}}, \bibinfo {author} {\bibfnamefont
  {Y.}~\bibnamefont {Singh}}, \bibinfo {author} {\bibfnamefont
  {P.}~\bibnamefont {Gegenwart}}, \ and\ \bibinfo {author} {\bibfnamefont
  {J.~P.}\ \bibnamefont {Hill}},\ }\href {\doibase 10.1103/PhysRevB.83.220403}
  {\bibfield  {journal} {\bibinfo  {journal} {Phys. Rev. B}\ }\textbf {\bibinfo
  {volume} {83}},\ \bibinfo {pages} {220403} (\bibinfo {year}
  {2011})}\BibitemShut {NoStop}%
\bibitem [{\citenamefont {Choi}\ \emph {et~al.}(2012)\citenamefont {Choi},
  \citenamefont {Coldea}, \citenamefont {Kolmogorov}, \citenamefont
  {Lancaster}, \citenamefont {Mazin}, \citenamefont {Blundell}, \citenamefont
  {Radaelli}, \citenamefont {Singh}, \citenamefont {Gegenwart}, \citenamefont
  {Choi}, \citenamefont {Cheong}, \citenamefont {Baker}, \citenamefont
  {Stock},\ and\ \citenamefont {Taylor}}]{Choi2012}%
  \BibitemOpen
  \bibfield  {author} {\bibinfo {author} {\bibfnamefont {S.~K.}\ \bibnamefont
  {Choi}}, \bibinfo {author} {\bibfnamefont {R.}~\bibnamefont {Coldea}},
  \bibinfo {author} {\bibfnamefont {A.~N.}\ \bibnamefont {Kolmogorov}},
  \bibinfo {author} {\bibfnamefont {T.}~\bibnamefont {Lancaster}}, \bibinfo
  {author} {\bibfnamefont {I.~I.}\ \bibnamefont {Mazin}}, \bibinfo {author}
  {\bibfnamefont {S.~J.}\ \bibnamefont {Blundell}}, \bibinfo {author}
  {\bibfnamefont {P.~G.}\ \bibnamefont {Radaelli}}, \bibinfo {author}
  {\bibfnamefont {Y.}~\bibnamefont {Singh}}, \bibinfo {author} {\bibfnamefont
  {P.}~\bibnamefont {Gegenwart}}, \bibinfo {author} {\bibfnamefont {K.~R.}\
  \bibnamefont {Choi}}, \bibinfo {author} {\bibfnamefont {S.-W.}\ \bibnamefont
  {Cheong}}, \bibinfo {author} {\bibfnamefont {P.~J.}\ \bibnamefont {Baker}},
  \bibinfo {author} {\bibfnamefont {C.}~\bibnamefont {Stock}}, \ and\ \bibinfo
  {author} {\bibfnamefont {J.}~\bibnamefont {Taylor}},\ }\href {\doibase
  10.1103/PhysRevLett.108.127204} {\bibfield  {journal} {\bibinfo  {journal}
  {Phys. Rev. Lett.}\ }\textbf {\bibinfo {volume} {108}},\ \bibinfo {pages}
  {127204} (\bibinfo {year} {2012})}\BibitemShut {NoStop}%
\bibitem [{\citenamefont {Ye}\ \emph {et~al.}(2012)\citenamefont {Ye},
  \citenamefont {Chi}, \citenamefont {Cao}, \citenamefont {Chakoumakos},
  \citenamefont {Fernandez-Baca}, \citenamefont {Custelcean}, \citenamefont
  {Qi}, \citenamefont {Korneta},\ and\ \citenamefont {Cao}}]{Ye2012}%
  \BibitemOpen
  \bibfield  {author} {\bibinfo {author} {\bibfnamefont {F.}~\bibnamefont
  {Ye}}, \bibinfo {author} {\bibfnamefont {S.}~\bibnamefont {Chi}}, \bibinfo
  {author} {\bibfnamefont {H.}~\bibnamefont {Cao}}, \bibinfo {author}
  {\bibfnamefont {B.~C.}\ \bibnamefont {Chakoumakos}}, \bibinfo {author}
  {\bibfnamefont {J.~A.}\ \bibnamefont {Fernandez-Baca}}, \bibinfo {author}
  {\bibfnamefont {R.}~\bibnamefont {Custelcean}}, \bibinfo {author}
  {\bibfnamefont {T.~F.}\ \bibnamefont {Qi}}, \bibinfo {author} {\bibfnamefont
  {O.~B.}\ \bibnamefont {Korneta}}, \ and\ \bibinfo {author} {\bibfnamefont
  {G.}~\bibnamefont {Cao}},\ }\href {\doibase 10.1103/PhysRevB.85.180403}
  {\bibfield  {journal} {\bibinfo  {journal} {Phys. Rev. B}\ }\textbf {\bibinfo
  {volume} {85}},\ \bibinfo {pages} {180403} (\bibinfo {year}
  {2012})}\BibitemShut {NoStop}%
\bibitem [{\citenamefont {Sears}\ \emph {et~al.}(2015)\citenamefont {Sears},
  \citenamefont {Songvilay}, \citenamefont {Plumb}, \citenamefont {Clancy},
  \citenamefont {Qiu}, \citenamefont {Zhao}, \citenamefont {Parshall},\ and\
  \citenamefont {Kim}}]{Sears2015}%
  \BibitemOpen
  \bibfield  {author} {\bibinfo {author} {\bibfnamefont {J.~A.}\ \bibnamefont
  {Sears}}, \bibinfo {author} {\bibfnamefont {M.}~\bibnamefont {Songvilay}},
  \bibinfo {author} {\bibfnamefont {K.~W.}\ \bibnamefont {Plumb}}, \bibinfo
  {author} {\bibfnamefont {J.~P.}\ \bibnamefont {Clancy}}, \bibinfo {author}
  {\bibfnamefont {Y.}~\bibnamefont {Qiu}}, \bibinfo {author} {\bibfnamefont
  {Y.}~\bibnamefont {Zhao}}, \bibinfo {author} {\bibfnamefont {D.}~\bibnamefont
  {Parshall}}, \ and\ \bibinfo {author} {\bibfnamefont {Y.-J.}\ \bibnamefont
  {Kim}},\ }\href {\doibase 10.1103/PhysRevB.91.144420} {\bibfield  {journal}
  {\bibinfo  {journal} {Phys. Rev. B}\ }\textbf {\bibinfo {volume} {91}},\
  \bibinfo {pages} {144420} (\bibinfo {year} {2015})}\BibitemShut {NoStop}%
\bibitem [{\citenamefont {Johnson}\ \emph {et~al.}(2015)\citenamefont
  {Johnson}, \citenamefont {Williams}, \citenamefont {Haghighirad},
  \citenamefont {Singleton}, \citenamefont {Zapf}, \citenamefont {Manuel},
  \citenamefont {Mazin}, \citenamefont {Li}, \citenamefont {Jeschke},
  \citenamefont {Valent{\'{i}}},\ and\ \citenamefont {Coldea}}]{Johnson2015}%
  \BibitemOpen
  \bibfield  {author} {\bibinfo {author} {\bibfnamefont {R.~D.}\ \bibnamefont
  {Johnson}}, \bibinfo {author} {\bibfnamefont {S.~C.}\ \bibnamefont
  {Williams}}, \bibinfo {author} {\bibfnamefont {A.~A.}\ \bibnamefont
  {Haghighirad}}, \bibinfo {author} {\bibfnamefont {J.}~\bibnamefont
  {Singleton}}, \bibinfo {author} {\bibfnamefont {V.}~\bibnamefont {Zapf}},
  \bibinfo {author} {\bibfnamefont {P.}~\bibnamefont {Manuel}}, \bibinfo
  {author} {\bibfnamefont {I.~I.}\ \bibnamefont {Mazin}}, \bibinfo {author}
  {\bibfnamefont {Y.}~\bibnamefont {Li}}, \bibinfo {author} {\bibfnamefont
  {H.~O.}\ \bibnamefont {Jeschke}}, \bibinfo {author} {\bibfnamefont
  {R.}~\bibnamefont {Valent{\'{i}}}}, \ and\ \bibinfo {author} {\bibfnamefont
  {R.}~\bibnamefont {Coldea}},\ }\href {\doibase 10.1103/PhysRevB.92.235119}
  {\bibfield  {journal} {\bibinfo  {journal} {Phys. Rev. B}\ }\textbf {\bibinfo
  {volume} {92}},\ \bibinfo {pages} {235119} (\bibinfo {year}
  {2015})}\BibitemShut {NoStop}%
\bibitem [{\citenamefont {Williams}\ \emph {et~al.}(2016)\citenamefont
  {Williams}, \citenamefont {Johnson}, \citenamefont {Freund}, \citenamefont
  {Choi}, \citenamefont {Jesche}, \citenamefont {Kimchi}, \citenamefont
  {Manni}, \citenamefont {Bombardi}, \citenamefont {Manuel}, \citenamefont
  {Gegenwart},\ and\ \citenamefont {Coldea}}]{Williams2016}%
  \BibitemOpen
  \bibfield  {author} {\bibinfo {author} {\bibfnamefont {S.~C.}\ \bibnamefont
  {Williams}}, \bibinfo {author} {\bibfnamefont {R.~D.}\ \bibnamefont
  {Johnson}}, \bibinfo {author} {\bibfnamefont {F.}~\bibnamefont {Freund}},
  \bibinfo {author} {\bibfnamefont {S.}~\bibnamefont {Choi}}, \bibinfo {author}
  {\bibfnamefont {A.}~\bibnamefont {Jesche}}, \bibinfo {author} {\bibfnamefont
  {I.}~\bibnamefont {Kimchi}}, \bibinfo {author} {\bibfnamefont
  {S.}~\bibnamefont {Manni}}, \bibinfo {author} {\bibfnamefont
  {A.}~\bibnamefont {Bombardi}}, \bibinfo {author} {\bibfnamefont
  {P.}~\bibnamefont {Manuel}}, \bibinfo {author} {\bibfnamefont
  {P.}~\bibnamefont {Gegenwart}}, \ and\ \bibinfo {author} {\bibfnamefont
  {R.}~\bibnamefont {Coldea}},\ }\href {\doibase 10.1103/PhysRevB.93.195158}
  {\bibfield  {journal} {\bibinfo  {journal} {Phys. Rev. B}\ }\textbf {\bibinfo
  {volume} {93}},\ \bibinfo {pages} {195158} (\bibinfo {year}
  {2016})}\BibitemShut {NoStop}%
\bibitem [{\citenamefont {Kimchi}\ and\ \citenamefont
  {You}(2011)}]{Kimchi2011}%
  \BibitemOpen
  \bibfield  {author} {\bibinfo {author} {\bibfnamefont {I.}~\bibnamefont
  {Kimchi}}\ and\ \bibinfo {author} {\bibfnamefont {Y.-Z.}\ \bibnamefont
  {You}},\ }\href {\doibase 10.1103/PhysRevB.84.180407} {\bibfield  {journal}
  {\bibinfo  {journal} {Phys. Rev. B}\ }\textbf {\bibinfo {volume} {84}},\
  \bibinfo {pages} {180407} (\bibinfo {year} {2011})}\BibitemShut {NoStop}%
\bibitem [{\citenamefont {Rau}\ \emph {et~al.}(2014)\citenamefont {Rau},
  \citenamefont {Lee},\ and\ \citenamefont {Kee}}]{Rau2014a}%
  \BibitemOpen
  \bibfield  {author} {\bibinfo {author} {\bibfnamefont {J.~G.}\ \bibnamefont
  {Rau}}, \bibinfo {author} {\bibfnamefont {E.~K.-H.}\ \bibnamefont {Lee}}, \
  and\ \bibinfo {author} {\bibfnamefont {H.-Y.}\ \bibnamefont {Kee}},\ }\href
  {\doibase 10.1103/PhysRevLett.112.077204} {\bibfield  {journal} {\bibinfo
  {journal} {Phys. Rev. Lett.}\ }\textbf {\bibinfo {volume} {112}},\ \bibinfo
  {pages} {077204} (\bibinfo {year} {2014})}\BibitemShut {NoStop}%
\bibitem [{\citenamefont {Kitagawa}\ \emph {et~al.}(2018)\citenamefont
  {Kitagawa}, \citenamefont {Takayama}, \citenamefont {Matsumoto},
  \citenamefont {Kato}, \citenamefont {Takano}, \citenamefont {Kishimoto},
  \citenamefont {Bette}, \citenamefont {Dinnebier}, \citenamefont {Jackeli},\
  and\ \citenamefont {Takagi}}]{Kitagawa2018}%
  \BibitemOpen
  \bibfield  {author} {\bibinfo {author} {\bibfnamefont {K.}~\bibnamefont
  {Kitagawa}}, \bibinfo {author} {\bibfnamefont {T.}~\bibnamefont {Takayama}},
  \bibinfo {author} {\bibfnamefont {Y.}~\bibnamefont {Matsumoto}}, \bibinfo
  {author} {\bibfnamefont {A.}~\bibnamefont {Kato}}, \bibinfo {author}
  {\bibfnamefont {R.}~\bibnamefont {Takano}}, \bibinfo {author} {\bibfnamefont
  {Y.}~\bibnamefont {Kishimoto}}, \bibinfo {author} {\bibfnamefont
  {S.}~\bibnamefont {Bette}}, \bibinfo {author} {\bibfnamefont
  {R.}~\bibnamefont {Dinnebier}}, \bibinfo {author} {\bibfnamefont
  {G.}~\bibnamefont {Jackeli}}, \ and\ \bibinfo {author} {\bibfnamefont
  {H.}~\bibnamefont {Takagi}},\ }\href {\doibase 10.1038/nature25482}
  {\bibfield  {journal} {\bibinfo  {journal} {Nature}\ }\textbf {\bibinfo
  {volume} {554}},\ \bibinfo {pages} {341} (\bibinfo {year}
  {2018})}\BibitemShut {NoStop}%
\bibitem [{\citenamefont {Bl{\"{o}}chl}(1994)}]{Blochl1994}%
  \BibitemOpen
  \bibfield  {author} {\bibinfo {author} {\bibfnamefont {P.~E.}\ \bibnamefont
  {Bl{\"{o}}chl}},\ }\href {\doibase 10.1103/PhysRevB.50.17953} {\bibfield
  {journal} {\bibinfo  {journal} {Phys. Rev. B}\ }\textbf {\bibinfo {volume}
  {50}},\ \bibinfo {pages} {17953} (\bibinfo {year} {1994})}\BibitemShut
  {NoStop}%
\bibitem [{\citenamefont {Kresse}\ and\ \citenamefont
  {Joubert}(1999)}]{Kresse1999}%
  \BibitemOpen
  \bibfield  {author} {\bibinfo {author} {\bibfnamefont {G.}~\bibnamefont
  {Kresse}}\ and\ \bibinfo {author} {\bibfnamefont {D.}~\bibnamefont
  {Joubert}},\ }\href {\doibase 10.1103/PhysRevB.59.1758} {\bibfield  {journal}
  {\bibinfo  {journal} {Phys. Rev. B}\ }\textbf {\bibinfo {volume} {59}},\
  \bibinfo {pages} {1758} (\bibinfo {year} {1999})}\BibitemShut {NoStop}%
\bibitem [{\citenamefont {Perdew}\ \emph {et~al.}(1996)\citenamefont {Perdew},
  \citenamefont {Burke},\ and\ \citenamefont {Ernzerhof}}]{Perdew1996}%
  \BibitemOpen
  \bibfield  {author} {\bibinfo {author} {\bibfnamefont {J.~P.}\ \bibnamefont
  {Perdew}}, \bibinfo {author} {\bibfnamefont {K.}~\bibnamefont {Burke}}, \
  and\ \bibinfo {author} {\bibfnamefont {M.}~\bibnamefont {Ernzerhof}},\ }\href
  {\doibase 10.1103/PhysRevLett.77.3865} {\bibfield  {journal} {\bibinfo
  {journal} {Phys. Rev. Lett.}\ }\textbf {\bibinfo {volume} {77}},\ \bibinfo
  {pages} {3865} (\bibinfo {year} {1996})}\BibitemShut {NoStop}%
\bibitem [{\citenamefont {Kresse}\ and\ \citenamefont
  {Furthm{\"{u}}ller}(1996{\natexlab{a}})}]{Kresse1996}%
  \BibitemOpen
  \bibfield  {author} {\bibinfo {author} {\bibfnamefont {G.}~\bibnamefont
  {Kresse}}\ and\ \bibinfo {author} {\bibfnamefont {J.}~\bibnamefont
  {Furthm{\"{u}}ller}},\ }\href {\doibase 10.1016/0927-0256(96)00008-0}
  {\bibfield  {journal} {\bibinfo  {journal} {Comput. Mater. Sci.}\ }\textbf
  {\bibinfo {volume} {6}},\ \bibinfo {pages} {15} (\bibinfo {year}
  {1996}{\natexlab{a}})}\BibitemShut {NoStop}%
\bibitem [{\citenamefont {Kresse}\ and\ \citenamefont
  {Furthm{\"{u}}ller}(1996{\natexlab{b}})}]{Kresse1996a}%
  \BibitemOpen
  \bibfield  {author} {\bibinfo {author} {\bibfnamefont {G.}~\bibnamefont
  {Kresse}}\ and\ \bibinfo {author} {\bibfnamefont {J.}~\bibnamefont
  {Furthm{\"{u}}ller}},\ }\href {\doibase 10.1103/PhysRevB.54.11169} {\bibfield
   {journal} {\bibinfo  {journal} {Phys. Rev. B}\ }\textbf {\bibinfo {volume}
  {54}},\ \bibinfo {pages} {11169} (\bibinfo {year}
  {1996}{\natexlab{b}})}\BibitemShut {NoStop}%
\bibitem [{\citenamefont {Bette}\ \emph {et~al.}(2017)\citenamefont {Bette},
  \citenamefont {Takayama}, \citenamefont {Kitagawa}, \citenamefont {Takano},
  \citenamefont {Takagi},\ and\ \citenamefont {Dinnebier}}]{Bette2017}%
  \BibitemOpen
  \bibfield  {author} {\bibinfo {author} {\bibfnamefont {S.}~\bibnamefont
  {Bette}}, \bibinfo {author} {\bibfnamefont {T.}~\bibnamefont {Takayama}},
  \bibinfo {author} {\bibfnamefont {K.}~\bibnamefont {Kitagawa}}, \bibinfo
  {author} {\bibfnamefont {R.}~\bibnamefont {Takano}}, \bibinfo {author}
  {\bibfnamefont {H.}~\bibnamefont {Takagi}}, \ and\ \bibinfo {author}
  {\bibfnamefont {R.~E.}\ \bibnamefont {Dinnebier}},\ }\href {\doibase
  10.1039/C7DT02978K} {\bibfield  {journal} {\bibinfo  {journal} {Dalt.
  Trans.}\ }\textbf {\bibinfo {volume} {46}},\ \bibinfo {pages} {15216}
  (\bibinfo {year} {2017})}\BibitemShut {NoStop}%
\bibitem [{\citenamefont {Wang}\ and\ \citenamefont {Xiang}(2014)}]{Wang2014f}%
  \BibitemOpen
  \bibfield  {author} {\bibinfo {author} {\bibfnamefont {P.~S.}\ \bibnamefont
  {Wang}}\ and\ \bibinfo {author} {\bibfnamefont {H.~J.}\ \bibnamefont
  {Xiang}},\ }\href {\doibase 10.1103/PhysRevX.4.011035} {\bibfield  {journal}
  {\bibinfo  {journal} {Phys. Rev. X}\ }\textbf {\bibinfo {volume} {4}},\
  \bibinfo {pages} {011035} (\bibinfo {year} {2014})}\BibitemShut {NoStop}%
\bibitem [{\citenamefont {Barrett}(1952)}]{Barrett1952}%
  \BibitemOpen
  \bibfield  {author} {\bibinfo {author} {\bibfnamefont {J.~H.}\ \bibnamefont
  {Barrett}},\ }\href {\doibase 10.1103/PhysRev.86.118} {\bibfield  {journal}
  {\bibinfo  {journal} {Phys. Rev.}\ }\textbf {\bibinfo {volume} {86}},\
  \bibinfo {pages} {118} (\bibinfo {year} {1952})}\BibitemShut {NoStop}%
\bibitem [{\citenamefont {Hu}\ \emph {et~al.}(2015)\citenamefont {Hu},
  \citenamefont {Wang},\ and\ \citenamefont {Feng}}]{Hu2015b}%
  \BibitemOpen
  \bibfield  {author} {\bibinfo {author} {\bibfnamefont {K.}~\bibnamefont
  {Hu}}, \bibinfo {author} {\bibfnamefont {F.}~\bibnamefont {Wang}}, \ and\
  \bibinfo {author} {\bibfnamefont {J.}~\bibnamefont {Feng}},\ }\href {\doibase
  10.1103/PhysRevLett.115.167204} {\bibfield  {journal} {\bibinfo  {journal}
  {Phys. Rev. Lett.}\ }\textbf {\bibinfo {volume} {115}},\ \bibinfo {pages}
  {167204} (\bibinfo {year} {2015})}\BibitemShut {NoStop}%
\bibitem [{\citenamefont {Winter}\ \emph {et~al.}(2016)\citenamefont {Winter},
  \citenamefont {Li}, \citenamefont {Jeschke},\ and\ \citenamefont
  {Valent{\'{i}}}}]{Winter2016}%
  \BibitemOpen
  \bibfield  {author} {\bibinfo {author} {\bibfnamefont {S.~M.}\ \bibnamefont
  {Winter}}, \bibinfo {author} {\bibfnamefont {Y.}~\bibnamefont {Li}}, \bibinfo
  {author} {\bibfnamefont {H.~O.}\ \bibnamefont {Jeschke}}, \ and\ \bibinfo
  {author} {\bibfnamefont {R.}~\bibnamefont {Valent{\'{i}}}},\ }\href {\doibase
  10.1103/PhysRevB.93.214431} {\bibfield  {journal} {\bibinfo  {journal} {Phys.
  Rev. B}\ }\textbf {\bibinfo {volume} {93}},\ \bibinfo {pages} {214431}
  (\bibinfo {year} {2016})}\BibitemShut {NoStop}%
\bibitem [{\citenamefont {Li}\ \emph {et~al.}(2018)\citenamefont {Li},
  \citenamefont {Winter},\ and\ \citenamefont {Valent{\'{i}}}}]{Li2018a}%
  \BibitemOpen
  \bibfield  {author} {\bibinfo {author} {\bibfnamefont {Y.}~\bibnamefont
  {Li}}, \bibinfo {author} {\bibfnamefont {S.~M.}\ \bibnamefont {Winter}}, \
  and\ \bibinfo {author} {\bibfnamefont {R.}~\bibnamefont {Valent{\'{i}}}},\
  }\href {\doibase 10.1103/PhysRevLett.121.247202} {\bibfield  {journal}
  {\bibinfo  {journal} {Phys. Rev. Lett.}\ }\textbf {\bibinfo {volume} {121}},\
  \bibinfo {pages} {247202} (\bibinfo {year} {2018})}\BibitemShut {NoStop}%
\bibitem [{\citenamefont {Yadav}\ \emph {et~al.}(2018)\citenamefont {Yadav},
  \citenamefont {Ray}, \citenamefont {Eldeeb}, \citenamefont {Nishimoto},
  \citenamefont {Hozoi},\ and\ \citenamefont {van~den Brink}}]{Yadav2018}%
  \BibitemOpen
  \bibfield  {author} {\bibinfo {author} {\bibfnamefont {R.}~\bibnamefont
  {Yadav}}, \bibinfo {author} {\bibfnamefont {R.}~\bibnamefont {Ray}}, \bibinfo
  {author} {\bibfnamefont {M.~S.}\ \bibnamefont {Eldeeb}}, \bibinfo {author}
  {\bibfnamefont {S.}~\bibnamefont {Nishimoto}}, \bibinfo {author}
  {\bibfnamefont {L.}~\bibnamefont {Hozoi}}, \ and\ \bibinfo {author}
  {\bibfnamefont {J.}~\bibnamefont {van~den Brink}},\ }\href {\doibase
  10.1103/PhysRevLett.121.197203} {\bibfield  {journal} {\bibinfo  {journal}
  {Phys. Rev. Lett.}\ }\textbf {\bibinfo {volume} {121}},\ \bibinfo {pages}
  {197203} (\bibinfo {year} {2018})}\BibitemShut {NoStop}%
\bibitem [{\citenamefont {Geirhos}\ \emph {et~al.}()\citenamefont {Geirhos},
  \citenamefont {Lunkenheimer}, \citenamefont {Blankenhorn}, \citenamefont
  {Claus}, \citenamefont {Matsumoto}, \citenamefont {Kitagawa}, \citenamefont
  {Takayama}, \citenamefont {Takagi}, \citenamefont {K{\'{e}}zsm{\'{a}}rki},\
  and\ \citenamefont {Loidl}}]{Geirhos2020}%
  \BibitemOpen
  \bibfield  {author} {\bibinfo {author} {\bibfnamefont {K.}~\bibnamefont
  {Geirhos}}, \bibinfo {author} {\bibfnamefont {P.}~\bibnamefont
  {Lunkenheimer}}, \bibinfo {author} {\bibfnamefont {M.}~\bibnamefont
  {Blankenhorn}}, \bibinfo {author} {\bibfnamefont {R.}~\bibnamefont {Claus}},
  \bibinfo {author} {\bibfnamefont {Y.}~\bibnamefont {Matsumoto}}, \bibinfo
  {author} {\bibfnamefont {K.}~\bibnamefont {Kitagawa}}, \bibinfo {author}
  {\bibfnamefont {T.}~\bibnamefont {Takayama}}, \bibinfo {author}
  {\bibfnamefont {H.}~\bibnamefont {Takagi}}, \bibinfo {author} {\bibfnamefont
  {I.}~\bibnamefont {K{\'{e}}zsm{\'{a}}rki}}, \ and\ \bibinfo {author}
  {\bibfnamefont {A.}~\bibnamefont {Loidl}},\ }\href
  {http://arxiv.org/abs/2002.09016} {\ }\Eprint
  {http://arxiv.org/abs/2002.09016} {arXiv:2002.09016} \BibitemShut {NoStop}%
\bibitem [{\citenamefont {Shen}\ \emph {et~al.}(2016)\citenamefont {Shen},
  \citenamefont {Wu}, \citenamefont {Song}, \citenamefont {Sun}, \citenamefont
  {Yang}, \citenamefont {Chai}, \citenamefont {Shang}, \citenamefont {Wang},
  \citenamefont {Scott},\ and\ \citenamefont {Sun}}]{Shen2016a}%
  \BibitemOpen
  \bibfield  {author} {\bibinfo {author} {\bibfnamefont {S.-P.}\ \bibnamefont
  {Shen}}, \bibinfo {author} {\bibfnamefont {J.-C.}\ \bibnamefont {Wu}},
  \bibinfo {author} {\bibfnamefont {J.-D.}\ \bibnamefont {Song}}, \bibinfo
  {author} {\bibfnamefont {X.-F.}\ \bibnamefont {Sun}}, \bibinfo {author}
  {\bibfnamefont {Y.-F.}\ \bibnamefont {Yang}}, \bibinfo {author}
  {\bibfnamefont {Y.-S.}\ \bibnamefont {Chai}}, \bibinfo {author}
  {\bibfnamefont {D.-S.}\ \bibnamefont {Shang}}, \bibinfo {author}
  {\bibfnamefont {S.-G.}\ \bibnamefont {Wang}}, \bibinfo {author}
  {\bibfnamefont {J.~F.}\ \bibnamefont {Scott}}, \ and\ \bibinfo {author}
  {\bibfnamefont {Y.}~\bibnamefont {Sun}},\ }\href {\doibase
  10.1038/ncomms10569} {\bibfield  {journal} {\bibinfo  {journal} {Nat.
  Commun.}\ }\textbf {\bibinfo {volume} {7}},\ \bibinfo {pages} {10569}
  (\bibinfo {year} {2016})}\BibitemShut {NoStop}%
\bibitem [{\citenamefont {Shimozawa}\ \emph {et~al.}(2017)\citenamefont
  {Shimozawa}, \citenamefont {Hashimoto}, \citenamefont {Ueda}, \citenamefont
  {Suzuki}, \citenamefont {Sugii}, \citenamefont {Yamada}, \citenamefont
  {Imai}, \citenamefont {Kobayashi}, \citenamefont {Itoh}, \citenamefont
  {Iguchi}, \citenamefont {Naka}, \citenamefont {Ishihara}, \citenamefont
  {Mori}, \citenamefont {Sasaki},\ and\ \citenamefont
  {Yamashita}}]{Shimozawa2017}%
  \BibitemOpen
  \bibfield  {author} {\bibinfo {author} {\bibfnamefont {M.}~\bibnamefont
  {Shimozawa}}, \bibinfo {author} {\bibfnamefont {K.}~\bibnamefont
  {Hashimoto}}, \bibinfo {author} {\bibfnamefont {A.}~\bibnamefont {Ueda}},
  \bibinfo {author} {\bibfnamefont {Y.}~\bibnamefont {Suzuki}}, \bibinfo
  {author} {\bibfnamefont {K.}~\bibnamefont {Sugii}}, \bibinfo {author}
  {\bibfnamefont {S.}~\bibnamefont {Yamada}}, \bibinfo {author} {\bibfnamefont
  {Y.}~\bibnamefont {Imai}}, \bibinfo {author} {\bibfnamefont {R.}~\bibnamefont
  {Kobayashi}}, \bibinfo {author} {\bibfnamefont {K.}~\bibnamefont {Itoh}},
  \bibinfo {author} {\bibfnamefont {S.}~\bibnamefont {Iguchi}}, \bibinfo
  {author} {\bibfnamefont {M.}~\bibnamefont {Naka}}, \bibinfo {author}
  {\bibfnamefont {S.}~\bibnamefont {Ishihara}}, \bibinfo {author}
  {\bibfnamefont {H.}~\bibnamefont {Mori}}, \bibinfo {author} {\bibfnamefont
  {T.}~\bibnamefont {Sasaki}}, \ and\ \bibinfo {author} {\bibfnamefont
  {M.}~\bibnamefont {Yamashita}},\ }\href {\doibase 10.1038/s41467-017-01849-x}
  {\bibfield  {journal} {\bibinfo  {journal} {Nat. Commun.}\ }\textbf {\bibinfo
  {volume} {8}},\ \bibinfo {pages} {1821} (\bibinfo {year} {2017})}\BibitemShut
  {NoStop}%
\bibitem [{\citenamefont {Hassan}\ \emph {et~al.}(2018)\citenamefont {Hassan},
  \citenamefont {Cunningham}, \citenamefont {Mourigal}, \citenamefont
  {Zhilyaeva}, \citenamefont {Torunova}, \citenamefont {Lyubovskaya},
  \citenamefont {Schlueter},\ and\ \citenamefont {Drichko}}]{Hassan2018}%
  \BibitemOpen
  \bibfield  {author} {\bibinfo {author} {\bibfnamefont {N.}~\bibnamefont
  {Hassan}}, \bibinfo {author} {\bibfnamefont {S.}~\bibnamefont {Cunningham}},
  \bibinfo {author} {\bibfnamefont {M.}~\bibnamefont {Mourigal}}, \bibinfo
  {author} {\bibfnamefont {E.~I.}\ \bibnamefont {Zhilyaeva}}, \bibinfo {author}
  {\bibfnamefont {S.~A.}\ \bibnamefont {Torunova}}, \bibinfo {author}
  {\bibfnamefont {R.~N.}\ \bibnamefont {Lyubovskaya}}, \bibinfo {author}
  {\bibfnamefont {J.~A.}\ \bibnamefont {Schlueter}}, \ and\ \bibinfo {author}
  {\bibfnamefont {N.}~\bibnamefont {Drichko}},\ }\href {\doibase
  10.1126/science.aan6286} {\bibfield  {journal} {\bibinfo  {journal}
  {Science}\ }\textbf {\bibinfo {volume} {360}},\ \bibinfo {pages} {1101}
  (\bibinfo {year} {2018})}\BibitemShut {NoStop}%
\bibitem [{\citenamefont {Zheng}\ \emph {et~al.}(2018)\citenamefont {Zheng},
  \citenamefont {Cui}, \citenamefont {Li}, \citenamefont {Ran}, \citenamefont
  {Wen},\ and\ \citenamefont {Yu}}]{Zheng2018}%
  \BibitemOpen
  \bibfield  {author} {\bibinfo {author} {\bibfnamefont {J.}~\bibnamefont
  {Zheng}}, \bibinfo {author} {\bibfnamefont {Y.}~\bibnamefont {Cui}}, \bibinfo
  {author} {\bibfnamefont {T.}~\bibnamefont {Li}}, \bibinfo {author}
  {\bibfnamefont {K.}~\bibnamefont {Ran}}, \bibinfo {author} {\bibfnamefont
  {J.}~\bibnamefont {Wen}}, \ and\ \bibinfo {author} {\bibfnamefont
  {W.}~\bibnamefont {Yu}},\ }\href {\doibase 10.1007/s11433-017-9166-1}
  {\bibfield  {journal} {\bibinfo  {journal} {Sci. China Physics, Mech.
  Astron.}\ }\textbf {\bibinfo {volume} {61}},\ \bibinfo {pages} {057021}
  (\bibinfo {year} {2018})}\BibitemShut {NoStop}%
\bibitem [{\citenamefont {Slagle}\ \emph {et~al.}(2018)\citenamefont {Slagle},
  \citenamefont {Choi}, \citenamefont {Chern},\ and\ \citenamefont
  {Kim}}]{Slagle2018}%
  \BibitemOpen
  \bibfield  {author} {\bibinfo {author} {\bibfnamefont {K.}~\bibnamefont
  {Slagle}}, \bibinfo {author} {\bibfnamefont {W.}~\bibnamefont {Choi}},
  \bibinfo {author} {\bibfnamefont {L.~E.}\ \bibnamefont {Chern}}, \ and\
  \bibinfo {author} {\bibfnamefont {Y.~B.}\ \bibnamefont {Kim}},\ }\href
  {\doibase 10.1103/PhysRevB.97.115159} {\bibfield  {journal} {\bibinfo
  {journal} {Phys. Rev. B}\ }\textbf {\bibinfo {volume} {97}},\ \bibinfo
  {pages} {115159} (\bibinfo {year} {2018})}\BibitemShut {NoStop}%
\bibitem [{\citenamefont {Kimchi}\ \emph {et~al.}(2018)\citenamefont {Kimchi},
  \citenamefont {Sheckelton}, \citenamefont {McQueen},\ and\ \citenamefont
  {Lee}}]{Kimchi2018}%
  \BibitemOpen
  \bibfield  {author} {\bibinfo {author} {\bibfnamefont {I.}~\bibnamefont
  {Kimchi}}, \bibinfo {author} {\bibfnamefont {J.~P.}\ \bibnamefont
  {Sheckelton}}, \bibinfo {author} {\bibfnamefont {T.~M.}\ \bibnamefont
  {McQueen}}, \ and\ \bibinfo {author} {\bibfnamefont {P.~A.}\ \bibnamefont
  {Lee}},\ }\href {\doibase 10.1038/s41467-018-06800-2} {\bibfield  {journal}
  {\bibinfo  {journal} {Nat. Commun.}\ }\textbf {\bibinfo {volume} {9}},\
  \bibinfo {pages} {4367} (\bibinfo {year} {2018})}\BibitemShut {NoStop}%
\end{thebibliography}%


\begin{thebibliography}{15}%
\makeatletter
\providecommand \@ifxundefined [1]{%
 \@ifx{#1\undefined}
}%
\providecommand \@ifnum [1]{%
 \ifnum #1\expandafter \@firstoftwo
 \else \expandafter \@secondoftwo
 \fi
}%
\providecommand \@ifx [1]{%
 \ifx #1\expandafter \@firstoftwo
 \else \expandafter \@secondoftwo
 \fi
}%
\providecommand \natexlab [1]{#1}%
\providecommand \enquote  [1]{``#1''}%
\providecommand \bibnamefont  [1]{#1}%
\providecommand \bibfnamefont [1]{#1}%
\providecommand \citenamefont [1]{#1}%
\providecommand \href@noop [0]{\@secondoftwo}%
\providecommand \href [0]{\begingroup \@sanitize@url \@href}%
\providecommand \@href[1]{\@@startlink{#1}\@@href}%
\providecommand \@@href[1]{\endgroup#1\@@endlink}%
\providecommand \@sanitize@url [0]{\catcode `\\12\catcode `\$12\catcode
  `\&12\catcode `\#12\catcode `\^12\catcode `\_12\catcode `\%12\relax}%
\providecommand \@@startlink[1]{}%
\providecommand \@@endlink[0]{}%
\providecommand \url  [0]{\begingroup\@sanitize@url \@url }%
\providecommand \@url [1]{\endgroup\@href {#1}{\urlprefix }}%
\providecommand \urlprefix  [0]{URL }%
\providecommand \Eprint [0]{\href }%
\providecommand \doibase [0]{http://dx.doi.org/}%
\providecommand \selectlanguage [0]{\@gobble}%
\providecommand \bibinfo  [0]{\@secondoftwo}%
\providecommand \bibfield  [0]{\@secondoftwo}%
\providecommand \translation [1]{[#1]}%
\providecommand \BibitemOpen [0]{}%
\providecommand \bibitemStop [0]{}%
\providecommand \bibitemNoStop [0]{.\EOS\space}%
\providecommand \EOS [0]{\spacefactor3000\relax}%
\providecommand \BibitemShut  [1]{\csname bibitem#1\endcsname}%
\let\auto@bib@innerbib\@empty
\bibitem [{\citenamefont {Bl{\"{o}}chl}(1994)}]{Blochl1994}%
  \BibitemOpen
  \bibfield  {author} {\bibinfo {author} {\bibfnamefont {P.~E.}\ \bibnamefont
  {Bl{\"{o}}chl}},\ }\href {\doibase 10.1103/PhysRevB.50.17953} {\bibfield
  {journal} {\bibinfo  {journal} {Phys. Rev. B}\ }\textbf {\bibinfo {volume}
  {50}},\ \bibinfo {pages} {17953} (\bibinfo {year} {1994})}\BibitemShut
  {NoStop}%
\bibitem [{\citenamefont {Kresse}\ and\ \citenamefont
  {Joubert}(1999)}]{Kresse1999}%
  \BibitemOpen
  \bibfield  {author} {\bibinfo {author} {\bibfnamefont {G.}~\bibnamefont
  {Kresse}}\ and\ \bibinfo {author} {\bibfnamefont {D.}~\bibnamefont
  {Joubert}},\ }\href {\doibase 10.1103/PhysRevB.59.1758} {\bibfield  {journal}
  {\bibinfo  {journal} {Phys. Rev. B}\ }\textbf {\bibinfo {volume} {59}},\
  \bibinfo {pages} {1758} (\bibinfo {year} {1999})}\BibitemShut {NoStop}%
\bibitem [{\citenamefont {Perdew}\ \emph {et~al.}(1996)\citenamefont {Perdew},
  \citenamefont {Burke},\ and\ \citenamefont {Ernzerhof}}]{Perdew1996}%
  \BibitemOpen
  \bibfield  {author} {\bibinfo {author} {\bibfnamefont {J.~P.}\ \bibnamefont
  {Perdew}}, \bibinfo {author} {\bibfnamefont {K.}~\bibnamefont {Burke}}, \
  and\ \bibinfo {author} {\bibfnamefont {M.}~\bibnamefont {Ernzerhof}},\ }\href
  {\doibase 10.1103/PhysRevLett.77.3865} {\bibfield  {journal} {\bibinfo
  {journal} {Phys. Rev. Lett.}\ }\textbf {\bibinfo {volume} {77}},\ \bibinfo
  {pages} {3865} (\bibinfo {year} {1996})}\BibitemShut {NoStop}%
\bibitem [{\citenamefont {Kresse}\ and\ \citenamefont
  {Furthm{\"{u}}ller}(1996{\natexlab{a}})}]{Kresse1996}%
  \BibitemOpen
  \bibfield  {author} {\bibinfo {author} {\bibfnamefont {G.}~\bibnamefont
  {Kresse}}\ and\ \bibinfo {author} {\bibfnamefont {J.}~\bibnamefont
  {Furthm{\"{u}}ller}},\ }\href {\doibase 10.1016/0927-0256(96)00008-0}
  {\bibfield  {journal} {\bibinfo  {journal} {Comput. Mater. Sci.}\ }\textbf
  {\bibinfo {volume} {6}},\ \bibinfo {pages} {15} (\bibinfo {year}
  {1996}{\natexlab{a}})}\BibitemShut {NoStop}%
\bibitem [{\citenamefont {Kresse}\ and\ \citenamefont
  {Furthm{\"{u}}ller}(1996{\natexlab{b}})}]{Kresse1996a}%
  \BibitemOpen
  \bibfield  {author} {\bibinfo {author} {\bibfnamefont {G.}~\bibnamefont
  {Kresse}}\ and\ \bibinfo {author} {\bibfnamefont {J.}~\bibnamefont
  {Furthm{\"{u}}ller}},\ }\href {\doibase 10.1103/PhysRevB.54.11169} {\bibfield
   {journal} {\bibinfo  {journal} {Phys. Rev. B}\ }\textbf {\bibinfo {volume}
  {54}},\ \bibinfo {pages} {11169} (\bibinfo {year}
  {1996}{\natexlab{b}})}\BibitemShut {NoStop}%
\bibitem [{\citenamefont {Monkhorst}\ and\ \citenamefont
  {Pack}(1976)}]{Monkhorst1976}%
  \BibitemOpen
  \bibfield  {author} {\bibinfo {author} {\bibfnamefont {H.~J.}\ \bibnamefont
  {Monkhorst}}\ and\ \bibinfo {author} {\bibfnamefont {J.~D.}\ \bibnamefont
  {Pack}},\ }\href {\doibase 10.1103/PhysRevB.13.5188} {\bibfield  {journal}
  {\bibinfo  {journal} {Phys. Rev. B}\ }\textbf {\bibinfo {volume} {13}},\
  \bibinfo {pages} {5188} (\bibinfo {year} {1976})}\BibitemShut {NoStop}%
\bibitem [{\citenamefont {Dudarev}\ \emph {et~al.}(1998)\citenamefont
  {Dudarev}, \citenamefont {Botton}, \citenamefont {Savrasov}, \citenamefont
  {Humphreys},\ and\ \citenamefont {Sutton}}]{Dudarev1998}%
  \BibitemOpen
  \bibfield  {author} {\bibinfo {author} {\bibfnamefont {S.~L.}\ \bibnamefont
  {Dudarev}}, \bibinfo {author} {\bibfnamefont {G.~A.}\ \bibnamefont {Botton}},
  \bibinfo {author} {\bibfnamefont {S.~Y.}\ \bibnamefont {Savrasov}}, \bibinfo
  {author} {\bibfnamefont {C.~J.}\ \bibnamefont {Humphreys}}, \ and\ \bibinfo
  {author} {\bibfnamefont {A.~P.}\ \bibnamefont {Sutton}},\ }\href {\doibase
  10.1103/PhysRevB.57.1505} {\bibfield  {journal} {\bibinfo  {journal} {Phys.
  Rev. B}\ }\textbf {\bibinfo {volume} {57}},\ \bibinfo {pages} {1505}
  (\bibinfo {year} {1998})}\BibitemShut {NoStop}%
\bibitem [{\citenamefont {Manni}\ \emph {et~al.}(2014)\citenamefont {Manni},
  \citenamefont {Choi}, \citenamefont {Mazin}, \citenamefont {Coldea},
  \citenamefont {Altmeyer}, \citenamefont {Jeschke}, \citenamefont
  {Valent{\'{i}}},\ and\ \citenamefont {Gegenwart}}]{Manni2014a}%
  \BibitemOpen
  \bibfield  {author} {\bibinfo {author} {\bibfnamefont {S.}~\bibnamefont
  {Manni}}, \bibinfo {author} {\bibfnamefont {S.}~\bibnamefont {Choi}},
  \bibinfo {author} {\bibfnamefont {I.~I.}\ \bibnamefont {Mazin}}, \bibinfo
  {author} {\bibfnamefont {R.}~\bibnamefont {Coldea}}, \bibinfo {author}
  {\bibfnamefont {M.}~\bibnamefont {Altmeyer}}, \bibinfo {author}
  {\bibfnamefont {H.~O.}\ \bibnamefont {Jeschke}}, \bibinfo {author}
  {\bibfnamefont {R.}~\bibnamefont {Valent{\'{i}}}}, \ and\ \bibinfo {author}
  {\bibfnamefont {P.}~\bibnamefont {Gegenwart}},\ }\href {\doibase
  10.1103/PhysRevB.89.245113} {\bibfield  {journal} {\bibinfo  {journal} {Phys.
  Rev. B}\ }\textbf {\bibinfo {volume} {89}},\ \bibinfo {pages} {245113}
  (\bibinfo {year} {2014})}\BibitemShut {NoStop}%
\bibitem [{\citenamefont {Hermann}\ \emph {et~al.}(2018)\citenamefont
  {Hermann}, \citenamefont {Altmeyer}, \citenamefont {Ebad-Allah},
  \citenamefont {Freund}, \citenamefont {Jesche}, \citenamefont {Tsirlin},
  \citenamefont {Hanfland}, \citenamefont {Gegenwart}, \citenamefont {Mazin},
  \citenamefont {Khomskii}, \citenamefont {Valent{\'{i}}},\ and\ \citenamefont
  {Kuntscher}}]{Hermann2018}%
  \BibitemOpen
  \bibfield  {author} {\bibinfo {author} {\bibfnamefont {V.}~\bibnamefont
  {Hermann}}, \bibinfo {author} {\bibfnamefont {M.}~\bibnamefont {Altmeyer}},
  \bibinfo {author} {\bibfnamefont {J.}~\bibnamefont {Ebad-Allah}}, \bibinfo
  {author} {\bibfnamefont {F.}~\bibnamefont {Freund}}, \bibinfo {author}
  {\bibfnamefont {A.}~\bibnamefont {Jesche}}, \bibinfo {author} {\bibfnamefont
  {A.~A.}\ \bibnamefont {Tsirlin}}, \bibinfo {author} {\bibfnamefont
  {M.}~\bibnamefont {Hanfland}}, \bibinfo {author} {\bibfnamefont
  {P.}~\bibnamefont {Gegenwart}}, \bibinfo {author} {\bibfnamefont {I.~I.}\
  \bibnamefont {Mazin}}, \bibinfo {author} {\bibfnamefont {D.~I.}\ \bibnamefont
  {Khomskii}}, \bibinfo {author} {\bibfnamefont {R.}~\bibnamefont
  {Valent{\'{i}}}}, \ and\ \bibinfo {author} {\bibfnamefont {C.~A.}\
  \bibnamefont {Kuntscher}},\ }\href {\doibase 10.1103/PhysRevB.97.020104}
  {\bibfield  {journal} {\bibinfo  {journal} {Phys. Rev. B}\ }\textbf {\bibinfo
  {volume} {97}},\ \bibinfo {pages} {020104} (\bibinfo {year}
  {2018})}\BibitemShut {NoStop}%
\bibitem [{\citenamefont {Bette}\ \emph {et~al.}(2017)\citenamefont {Bette},
  \citenamefont {Takayama}, \citenamefont {Kitagawa}, \citenamefont {Takano},
  \citenamefont {Takagi},\ and\ \citenamefont {Dinnebier}}]{Bette2017}%
  \BibitemOpen
  \bibfield  {author} {\bibinfo {author} {\bibfnamefont {S.}~\bibnamefont
  {Bette}}, \bibinfo {author} {\bibfnamefont {T.}~\bibnamefont {Takayama}},
  \bibinfo {author} {\bibfnamefont {K.}~\bibnamefont {Kitagawa}}, \bibinfo
  {author} {\bibfnamefont {R.}~\bibnamefont {Takano}}, \bibinfo {author}
  {\bibfnamefont {H.}~\bibnamefont {Takagi}}, \ and\ \bibinfo {author}
  {\bibfnamefont {R.~E.}\ \bibnamefont {Dinnebier}},\ }\href {\doibase
  10.1039/C7DT02978K} {\bibfield  {journal} {\bibinfo  {journal} {Dalt.
  Trans.}\ }\textbf {\bibinfo {volume} {46}},\ \bibinfo {pages} {15216}
  (\bibinfo {year} {2017})}\BibitemShut {NoStop}%
\bibitem [{\citenamefont {O'Malley}\ \emph {et~al.}(2008)\citenamefont
  {O'Malley}, \citenamefont {Verweij},\ and\ \citenamefont
  {Woodward}}]{OMalley2008}%
  \BibitemOpen
  \bibfield  {author} {\bibinfo {author} {\bibfnamefont {M.~J.}\ \bibnamefont
  {O'Malley}}, \bibinfo {author} {\bibfnamefont {H.}~\bibnamefont {Verweij}}, \
  and\ \bibinfo {author} {\bibfnamefont {P.~M.}\ \bibnamefont {Woodward}},\
  }\href {\doibase 10.1016/j.jssc.2008.04.005} {\bibfield  {journal} {\bibinfo
  {journal} {J. Solid State Chem.}\ }\textbf {\bibinfo {volume} {181}},\
  \bibinfo {pages} {1803} (\bibinfo {year} {2008})}\BibitemShut {NoStop}%
\bibitem [{\citenamefont {Hobbs}\ \emph {et~al.}(2000)\citenamefont {Hobbs},
  \citenamefont {Kresse},\ and\ \citenamefont {Hafner}}]{Hobbs2000}%
  \BibitemOpen
  \bibfield  {author} {\bibinfo {author} {\bibfnamefont {D.}~\bibnamefont
  {Hobbs}}, \bibinfo {author} {\bibfnamefont {G.}~\bibnamefont {Kresse}}, \
  and\ \bibinfo {author} {\bibfnamefont {J.}~\bibnamefont {Hafner}},\ }\href
  {\doibase 10.1103/PhysRevB.62.11556} {\bibfield  {journal} {\bibinfo
  {journal} {Phys. Rev. B}\ }\textbf {\bibinfo {volume} {62}},\ \bibinfo
  {pages} {11556} (\bibinfo {year} {2000})}\BibitemShut {NoStop}%
\bibitem [{\citenamefont {Hu}\ \emph {et~al.}(2015)\citenamefont {Hu},
  \citenamefont {Wang},\ and\ \citenamefont {Feng}}]{Hu2015b}%
  \BibitemOpen
  \bibfield  {author} {\bibinfo {author} {\bibfnamefont {K.}~\bibnamefont
  {Hu}}, \bibinfo {author} {\bibfnamefont {F.}~\bibnamefont {Wang}}, \ and\
  \bibinfo {author} {\bibfnamefont {J.}~\bibnamefont {Feng}},\ }\href {\doibase
  10.1103/PhysRevLett.115.167204} {\bibfield  {journal} {\bibinfo  {journal}
  {Phys. Rev. Lett.}\ }\textbf {\bibinfo {volume} {115}},\ \bibinfo {pages}
  {167204} (\bibinfo {year} {2015})}\BibitemShut {NoStop}%
\bibitem [{\citenamefont {Li}\ \emph {et~al.}(2018)\citenamefont {Li},
  \citenamefont {Winter},\ and\ \citenamefont {Valent{\'{i}}}}]{Li2018a}%
  \BibitemOpen
  \bibfield  {author} {\bibinfo {author} {\bibfnamefont {Y.}~\bibnamefont
  {Li}}, \bibinfo {author} {\bibfnamefont {S.~M.}\ \bibnamefont {Winter}}, \
  and\ \bibinfo {author} {\bibfnamefont {R.}~\bibnamefont {Valent{\'{i}}}},\
  }\href {\doibase 10.1103/PhysRevLett.121.247202} {\bibfield  {journal}
  {\bibinfo  {journal} {Phys. Rev. Lett.}\ }\textbf {\bibinfo {volume} {121}},\
  \bibinfo {pages} {247202} (\bibinfo {year} {2018})}\BibitemShut {NoStop}%
\bibitem [{\citenamefont {Gu}(2010)}]{Gu2010}%
  \BibitemOpen
  \bibfield  {author} {\bibinfo {author} {\bibfnamefont {S.-J.}\ \bibnamefont
  {Gu}},\ }\href {\doibase 10.1142/S0217979210056335} {\bibfield  {journal}
  {\bibinfo  {journal} {Int. J. Mod. Phys. B}\ }\textbf {\bibinfo {volume}
  {24}},\ \bibinfo {pages} {4371} (\bibinfo {year} {2010})}\BibitemShut
  {NoStop}%
\end{thebibliography}%
\end{document}


\title{Supplementary Materials for ``Possible Quantum Paraelectric State in Kitaev Spin Liquid Candidate H$_{3}$LiIr$_{2}$O$_{6}$''}
\author{Shuai Wang}
\affiliation{International Center for Quantum Materials, School of Physics, Peking University, Beijing 100871, China}
\author{Long Zhang}
\email{longzhang@ucas.ac.cn}
\affiliation{Kavli Institute for Theoretical Sciences and CAS Center for Excellence in Topological Quantum Computation, University of Chinese Academy of Sciences, Beijing 100190, China}
\author{Fa Wang}
\email{wangfa@pku.edu.cn}
\affiliation{International Center for Quantum Materials, School of Physics, Peking University, Beijing 100871, China}
\affiliation{Collaborative Innovation Center of Quantum Matter, Beijing 100871, China}
\date{\today}
\begin{abstract}
The supplementary materials include the implementation of first-principles calculations, the fitting of the electric dipole-dipole interactions, the calculations of the energy potentials of the protons, the calculations of the magnetic interactions, the band structures with different electric dipole configurations and an estimate of its impact on the magnetic interactions, and the exact diagonalization of the $K$-$J_{1}$-$J_{3}$ model.
\end{abstract}
\maketitle

\section{Implementation of first-principles calculations}

First-principles calculations are performed with the projector augmented wave (PAW) method \cite{Blochl1994, Kresse1999} and the generalized gradient approximation (GGA) to the exchange-correlation functional \cite{Perdew1996} implemented in the Vienna {\it ab initio} Simulation Package (VASP) \cite{Kresse1996, Kresse1996a}. The energy cutoff of plane waves is set to be 500 eV. The reciprocal lattice is sampled within a $9 \times 5\times 9 $ and a $11 \times 6 \times 11$ $\Gamma$-centered Monkhorst-Pack grid \cite{Monkhorst1976} in the structural optimization and the magnetic calculations, respectively. The spin-orbit coupling and the onsite Coulomb interaction on the iridium atoms are included in the noncollinear magnetic calculations. The effective Coulomb repulsion $U_{\mathrm{eff}}=U-J$ \cite{Dudarev1998} is set to be $1.5$ eV, the same as in the DFT calculations of $\alpha$-Li$_{2}$IrO$_{3}$ \cite{Manni2014a, Hermann2018}.

The lattice constants and the atom coordinates except those of the hydrogen atoms are taken from the experimental data \cite{Bette2017}. The hydrogen atom positions are fully relaxed and optimized in calculations. The atomic coordinates of the interlayer lithium atoms in $\alpha$-Li$_2$IrO$_3$ \cite{OMalley2008} are used as the initial positions of the hydrogen atoms in the structural optimization, because H$_3$LiIr$_2$O$_6$ was synthesized by substituting hydrogen for these interlayer lithium atoms. The structural parameters preserving the C2/m space group symmetry are listed in Table \ref{tab:structure}.

\section{Electric dipole-dipole interactions}

The total energies with different electric dipole configurations from the first-principles calculations are listed in Table \ref{tab:dipole}. Two different methods are adopted to extract the intraplane interactions $D_{1,2,3}$ and the interplane interactions $D'_{1,2,3}$ of the Ising model defined in Fig.~3 of the main text.

\begin{figure*}
\includegraphics[width=0.8\textwidth]{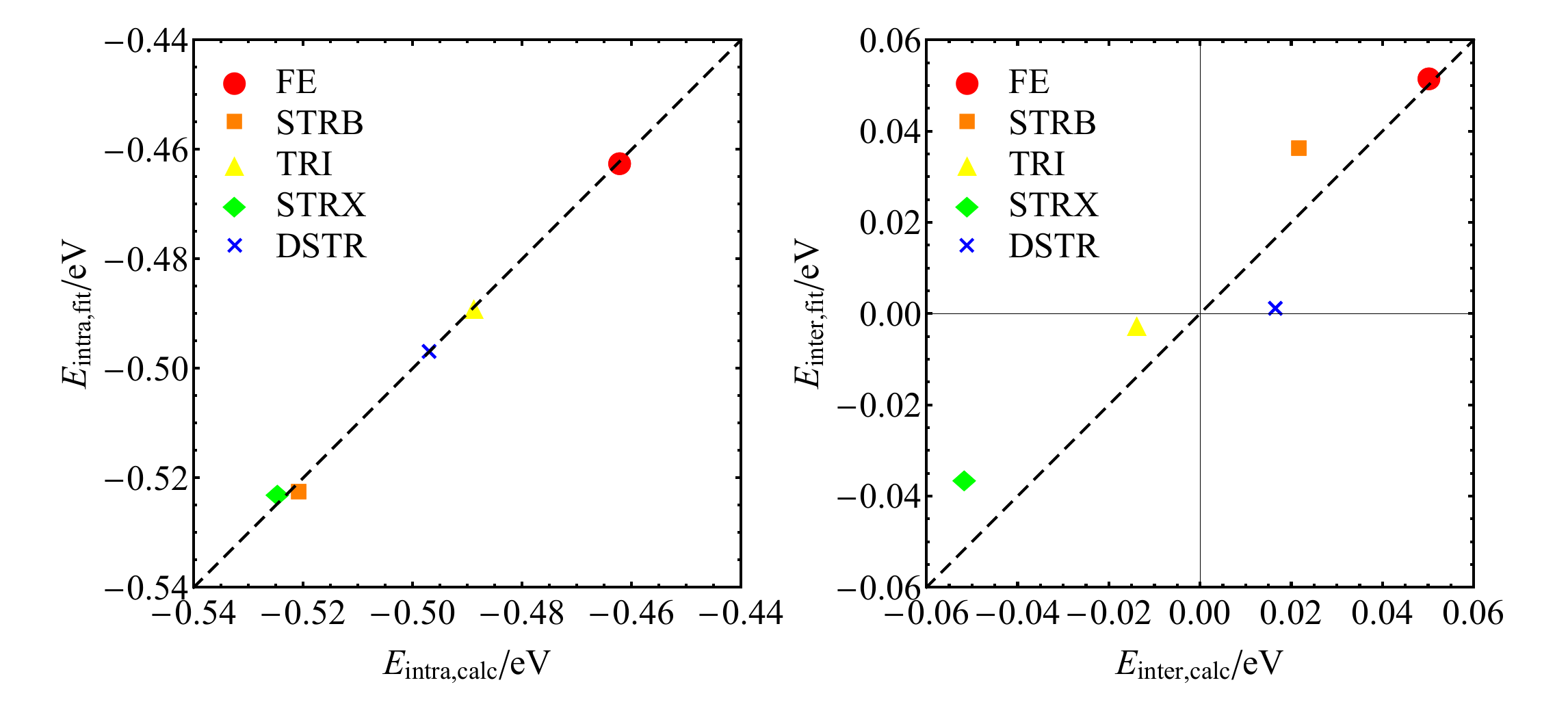}
\caption{(a) The intraplane interaction energies extracted from first-principles calculations vs. the fitted intraplane energies of the Ising model. (b) Similar comparison for the interplane energies. A constant -282.413 eV is subtracted in the intraplane interaction energies.}
\label{fig:inplane}
\end{figure*}

First, for a specific inplane dipole configuration, e.g., FE, its intraplane (interplane) interaction energy is the mean value (half of the difference) of the total energies with the same and the opposite interplane configurations, e.g., FE$+$FE and FE$-$FE. The intraplane and the interplane energies of several inplane dipole configurations are thus extracted and listed in Table \ref{tab:inplane}. The almost degenerate intraplane interaction energies of the STRB and the STRX configurations indicates that the spatial anisotropy in the dipole-dipole interaction is negligibly small. The intraplane and the interplane energies are fitted to the two sets of energies, respectively, and the least-square fitting yields $D_{1}=1.8$ meV, $D_{2}=-0.5$ meV, $D_{3}=-0.1$ meV, and $D'_{1}=-0.1$ meV, $D'_{2}=1.1$ meV, $D'_{3}=0.2$ meV. The comparison of the inplane and the interplane energies to the fitted results are shown in Fig.~\ref{fig:inplane}.

Second, we fit both the intraplane and the interplane interaction parameters directly to the total energies of twelve dipole configurations. The results are reported in the main text.

The interaction strengths derived with both methods are consistent with each other within 0.1--0.2 meV, thus suggesting that the Ising model correctly captures the electric dipole-dipole interactions in H$_{3}$LiIr$_{2}$O$_{6}$.

\section{Proton energy potentials and fluctuations}

The potential energy of each proton is calculated by varying its height along the O-H-O bond while fixing other protons at their bond centers in the first-principle calculations.

There are two types of O-H-O bonds, which are formed by O(2)-H(1)-O(2) and O(1)-H(2)-O(1), respectively. The inplane coordinates of the O(2)-H(1)-O(2) bonds are $(0.5,0)$ or $(0,0.5)$, and the O(2)-O(2) bond length is 2.54 {\AA}. The energy potential $V(z)$ calculated in a supercell with eight iridium atoms (forming two layers, with four in each layer) and the quantum tunneling strength $h_{x}$ of the H(1) ion have been reported in the main text. We also calculate the energy potential in a supercell with sixteen iridium atoms, and find consistent results within 1 meV with the smaller supercell (see Fig. \ref{fig:H2potential}), therefore the finite supercell size effect is negligible.

The inplane coordinates of the O(1)-H(2)-O(1) bonds are $(\pm 0.1808, 0)$ or $(0.5,\pm 0.3192)$, and the O(1)-O(1) bond length is 2.46 {\AA}. The energy potential of the H(2) ion is shown in Fig.~\ref{fig:H2potential}. The tunneling strength of H(2) are obtained from the energy difference of the bonding and the antibonding states, $h_{x}=62.2$ meV. It is even larger than that of H(1), thus also favors a quantum paraelectric ground state.

\section{Calculations of magnetic interactions}
 
The total energies of different magnetic configurations (Fig.~\ref{fig:mag_config}) are calculated with the spin moments constrained along the specified directions \cite{Hobbs2000}. The results are shown in Fig.~\ref{fig:magnetic}. The spin moment direction is varied in the crystallographic $ab$, $bc$, and $ac$ planes following a similar study on Na$_{2}$IrO$_{3}$ \cite{Hu2015b}. These energies are used to fit the interaction parameters in the (extended) Kitaev-Heisenberg model, Eq. (7) of the main text. Because the lengths of the $x$($y$)-bond and the $z$-bond are slightly different (see Fig.~1 of the main text), we adopt both a spatially isotropic model, and an anisotropic model that distinguishes the nearest $z$-bond parameters from the $x$- and $y$-bonds. The results are listed in Tables \ref{tab:KHmodel} and \ref{tab:anistropic}, respectively.

Moreover, we calculate and compare the total energies of the inplane zigzag order with ferromagnetic and antiferromagnetic interlayer stacking patterns (the spin moment is in the $ac$ plane, $\theta=110^{\circ}$). The energy difference is found to be less than $1$ meV per magnetic unit cell. Therefore, the interlayer magnetic interaction is weak and negligible.

\section{LDA band structures with different dipole configurations}

In order to estimate the impact of the electric dipole fluctuations on the magnetic interactions, we calculate the band structures with different electric dipole configurations. We adopt a supercell including two layers and four iridium atoms in each layer. The configurations of electric dipoles are denoted by FE$+$FE and FE$-$FE in Table \ref{tab:dipole}, i.e., the dipoles are ferroelectric within each layer, and are parallel (FE$+$FE) or antiparallel (FE$-$FE) to the other layer. In the FE$+$FE case, both layers of iridium atoms are equivalent, while in the FE$-$FE case, all protons move close to one of the iridium layers (referred to as ``hydrogen-rich'' in Ref. \cite{Li2018a}) and away from the other layer (``hydrogen-poor''). Therefore, the difference of their band structures reflect the impact of the dipole configurations.

The band structures are calculated with the local density approximation (LDA) without including spin-orbit coupling or correlation $U$, because they lead to long range magnetic order in the first-principle calculations, contradicting the experiments. Nevertheless, we believe that the strength of the electric dipole configuration effects can be qualitatively captured by the LDA calculations.

The LDA band structures along the high symmetry lines are shown in Fig.~\ref{fig:band}. The 24 bands near the Fermi energy mainly come from the iridium $t_{2g}$ levels. The energy bands in the FE$-$FE case are split and shifted compared with the FE$+$FE case. These changes are particularly pronounced in the $k_{z}=\pi$ plane in the folded Brillouin zone, i.e., along the Z-R-U-Z-T-R trajectory in Fig. \ref{fig:band}, which exactly captures the difference of the hydrogen-rich and the hydrogen-poor layers. The band splitting and shifting along this trajectory are about 0.1 eV, thus we estimate the change of the nearest neighbor hopping to be at the same order of magnitude, $\delta t\sim 0.1$ eV. With the onsite repulsion $U\sim 1.5$ eV, we can estimate the change of the magnetic interactions induced by the electric dipole fluctuations [Eq.~(9) in the main text], $\delta J\sim 5$ meV, which is comparable to our fitted magnetic interactions from the first-principle calculations. Therefore, the magnetic interactions will be strongly modified by the electric dipole fluctuations, and the ratio $J_{1}/|K_{1}|$ can be significantly reduced.

\section{Exact diagonalization of $K$-$J_{1}$-$J_{3}$ model}

We study the phase diagram of the $K$-$J_{1}$-$J_{3}$ model, 
$$
H=K\sum_{\langle ij\rangle\in \gamma}S_i^\gamma S_j^\gamma
+J_1\sum_{\langle ij\rangle}\vec{S}_i\cdot \vec{S}_j
+J_3\sum_{\langle\langle\langle ij\rangle\rangle\rangle}\vec{S}_i\cdot \vec{S}_j,
$$
by exact diagonalization on small lattices.
We set $K=-1$ because this is the sign of $K$ in H$_3$LiIr$_2$O$_6$ according to our first-principles calculation.

Two finite-size honeycomb lattices, $2\sqrt{3}\times 2\sqrt{3}\times 2$
with $24$ sites and $4\times 4\times 2$ with $32$ sites, 
are studied (see Fig.~\ref{fig:lattices}).
With periodic boundary conditions they both preserve the full lattice symmetries of the honeycomb lattice.

The ground state energies and wave functions of $H$ are obtained by the Lanczos method.
The following symmetries of $H$ are exploited to reduce the Hilbert space sizes, 
(a) lattice translations;
(b) conservation of $S_z$-parity, $\prod_j (2S_j^z)$; 
(c) a spin space 2-fold rotation $C_{2}^{\text{(spin }x\text{-axis)}}$ generated by unitary operator $\prod_j \exp(i\pi S_j^x)$; 
(d) a spatial 2-fold rotation $C_{2}^{\text{(}z\text{-bond)}}$ around a $z$-bond
combined with a spin space rotation by unitary operator 
$\prod_j \exp(i\frac{\pi}{2}S_j^z)$ in the translation trivial sector.
The ground states are found to be in the sector with 
trivial translations, and
$\prod_j (2S_j^z)=+1$, and
$C_{2}^{\text{(spin }x\text{-axis)}}$ eigenvalue $+1$, and
$C_{2}^{\text{(}z\text{-bond)}}$ eigenvalue $+1$.
The reduced Hilbert space sizes are $88763$ and $16798804$ for the $24$-site and $32$-site lattices respectively.
Phase boundaries are located by
the peaks of the ground state fidelity susceptibility \cite{Gu2010}, 
$\frac{2\cdot [1-|\langle \psi_0(\alpha+\delta\alpha)|\psi_0(\alpha)\rangle|]}{(\delta\alpha)^2}$ 
where $|\psi_0(\alpha)\rangle$ is the ground state under parameter $\alpha$.
The results for the two lattices are consistent (see
Fig.~5 in the main text).
Some detailed results for the $32$-site lattice are shown in Fig.~\ref{fig:phase-boundary}.

\bibliography{library,books}

\begin{ruledtabular}
\begin{table*}
\caption{The structural parameters of H$_3$LiIr$_2$O$_6$ with the space group C2/m. The lattice constants ($a = 5.3489$ {\AA}, $b=9.2431$ {\AA}, $c=4.8734$ {\AA}, $\beta=111.440^{\circ}$) and the atom positions except those of the hydrogen atoms are taken from the experimental data \cite{Bette2017}.}
\label{tab:structure}
\begin{tabular}{c|cccccc}
Atoms 	& Ir	& Li	& O(1)	& O(2)	& H(1)	& H(2)	\\
\hline
$x$		& 0		& 0		& 0.404	& 0.417	& 0		& 0		\\
$y$		& 0.335	& 0		& 0.323	& 0		& 0.5	& 0.8192\\
$z$		& 0		& 0		& 0.229	& 0.220	& 0.5	& 0.5
\end{tabular}
\end{table*}
\end{ruledtabular}

\begin{ruledtabular}
\begin{table*}
\centering
\caption{Ground state energies (in eV) per supercell of different electric dipole configurations. The supercell in calculations consists of twelve hydrogen atoms in two layers. The dots and the crosses in the schematic dipole configurations stand for the electric dipoles pointing upward or downward, respectively.}
\label{tab:dipole}
\begin{tabular}{c|cccccc}
Label & FE$+$FE & FE$-$FE & STRB$+$STRB & STRB$-$STRB & TRI$+$TRI & TRI$-$TRI \\
\hline
Configuration & 
\mgape{\includegraphics[width=0.08\textwidth]{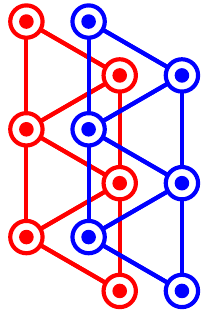}} &
\mgape{\includegraphics[width=0.08\textwidth]{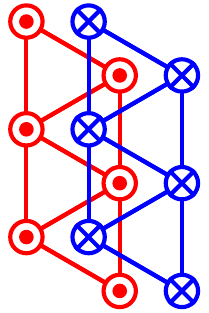}} &
\mgape{\includegraphics[width=0.08\textwidth]{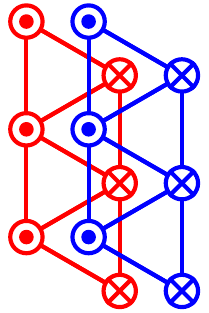}} &
\mgape{\includegraphics[width=0.08\textwidth]{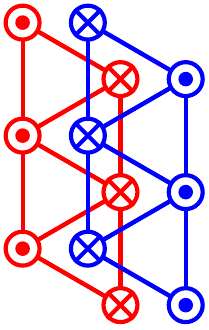}} &
\mgape{\includegraphics[width=0.08\textwidth]{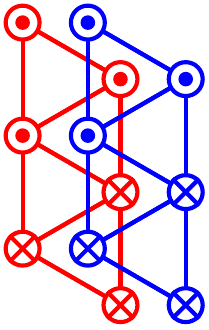}} &
\mgape{\includegraphics[width=0.08\textwidth]{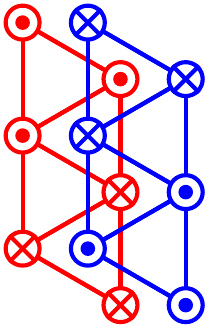}} \\
Energy & -282.8245 & -282.9252 & -282.9553 & -282.9112 & -282.9152 & -282.8877 \\
\hline\hline
Label & DSTR$+$DSTR & DSTR$-$DSTR & STRX$+$STRX & STRX$-$STRX & FE$+$STRB & FE$+$STRX \\
\hline
Configuration & 
\mgape{\includegraphics[width=0.08\textwidth]{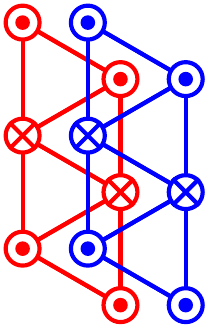}} &
\mgape{\includegraphics[width=0.08\textwidth]{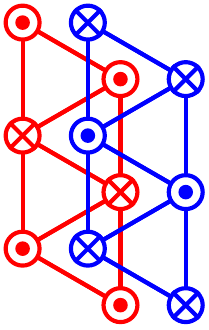}} &
\mgape{\includegraphics[width=0.117584\textwidth]{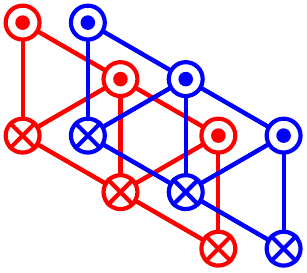}} &
\mgape{\includegraphics[width=0.117584\textwidth]{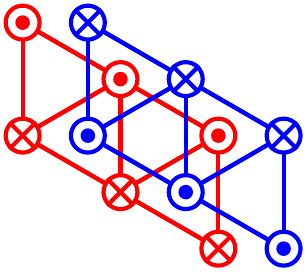}} &
\mgape{\includegraphics[width=0.08\textwidth]{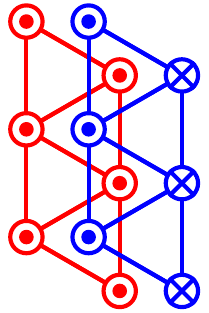}} &
\mgape{\includegraphics[width=0.117584\textwidth]{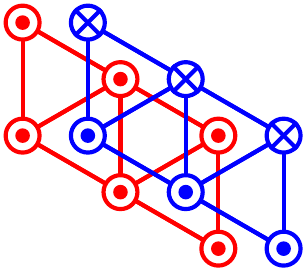}} \\
Energy & -282.8931 & -282.9261 & -282.9890 & -282.8857 &  -282.9070 & -282.9103 \\
\end{tabular}
\end{table*}
\end{ruledtabular}

\begin{ruledtabular}
\begin{table*}
\centering
\caption{The intraplane and the interplane dipole-dipole interaction energies, $E_{\mathrm{intra}}$ and $E_{\mathrm{inter}}$ (in eV) extracted from the total energies with the same and the opposite dipole configurations in the two layers.}
\label{tab:inplane}
\begin{tabular}{c|ccccc}
Label 					& FE 		& STRB 		& STRX 		& TRI 		& DSTR 		\\
\hline
$E_{\mathrm{intra}}$	& -282.875	& -282.934	& -282.937	& -282.901	& -282.910	\\
$E_{\mathrm{inter}}$	& 0.050		& 0.022		& -0.052	& -0.014	& 0.017		\\
\end{tabular}
\end{table*}
\end{ruledtabular}

\begin{ruledtabular}
\begin{table*}
\centering
\caption{The fitted magnetic interaction parameters (in meV/$g^2S^{2}$, $g$: the Land\'e factor, $S$: the size of the pseudospin) assuming the same interaction parameters on the $x$($y$)-bond and the $z$-bond. The uncertainty comes from the least-square fitting procedure.}
\label{tab:KHmodel}
\begin{tabular}{c|cccccc}
Parameter 	&	$K$			& $J_{1}$	& $J_{2}$	& $J_{3}$	& $\Gamma$	& $\Gamma'$	\\
\hline
Strength	& $-21.6(6)$	& $6.3(2)$	& $0.4(1)$	& $3.1(1)$	& $-0.2(3)$	& $-4.1(2)$	\\
\end{tabular}
\end{table*}
\end{ruledtabular}

\begin{ruledtabular}
\begin{table*}
\centering
\caption{The fitted magnetic interaction parameters (in meV/$g^2S^{2}$, $g$: the Land\'e factor, $S$: the size of the pseudospin) incorporating the inequivalence of the nearest $x$($y$)-bond and the $z$-bond. The uncertainty comes from the least-square fitting procedure.}
\label{tab:anistropic}
\begin{tabular}{c|cccccccccc}
Parameter 	& $K_{x(y)}$ & $K_{z}$ & $J_{x(y)}$ & $J_{z}$ & $\Gamma_{x(y)}$ & $\Gamma_{z}$ & $\Gamma^\prime_{x(y)}$ & $\Gamma^\prime_{z}$ & $J_2$ & $J_3$\\
\hline
Strength 	&$-24.1(6)$ & $-17.9(9)$ & $7.7(3)$ & $3.7(4)$ & $0.0(4)$ & $-0.3(3)$ & $-5.5(4)$ & $-4.0(2)$ & $0.4(1)$ & $3.3(1)$ \\
\end{tabular}
\end{table*}
\end{ruledtabular}

\begin{figure*}
\includegraphics[width=0.8\textwidth]{Hpotential}
\caption{(a) The 1D double-well potentials $V(z)$ of the H(1) ion along the O(2)-H(1)-O(2) bond direction calculated in two supercells including eight and sixteen iridium atoms, respectively. Both curves are shifted by a constant so that the zero energy points are at the bond center. (b) $V(z)$ of the H(2) ion along the O(1)-H(2)-O(1) bond direction. The proton wavefunctions of the bonding ($\Psi_{\mathrm{b}}$) and the antibonding ($\Psi_{\mathrm{ab}}$) states are also shown.}
\label{fig:H2potential}
\end{figure*}

\begin{figure*}
\centering
\includegraphics[width=0.8\textwidth]{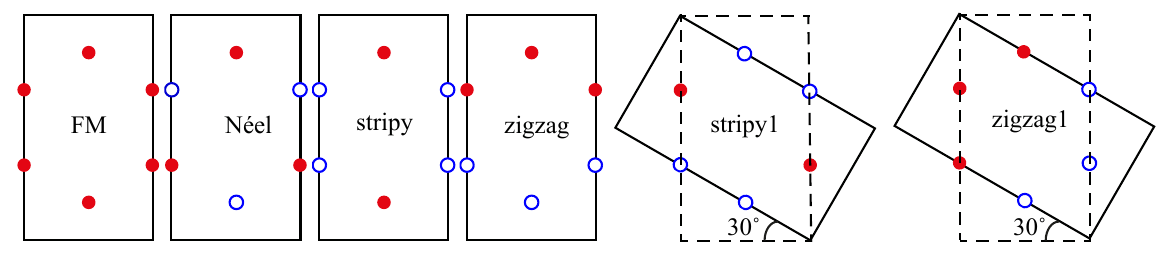}
\caption{Magnetic configuarations used in the magnetic calculations. The two types of spins, which are antiparallel to each other, are denoted by filled and empty circles, respectively. There are two inequivalent propagating directions of the stripy and the zigzag orders due to the bond length difference between the $x(y)$ bond and the $z$ bond. The solid lines enclose the magnetic unit cell used in the calculations.}
\label{fig:mag_config}
\end{figure*}

\begin{figure*}
\centering
\includegraphics[width=0.8\textwidth]{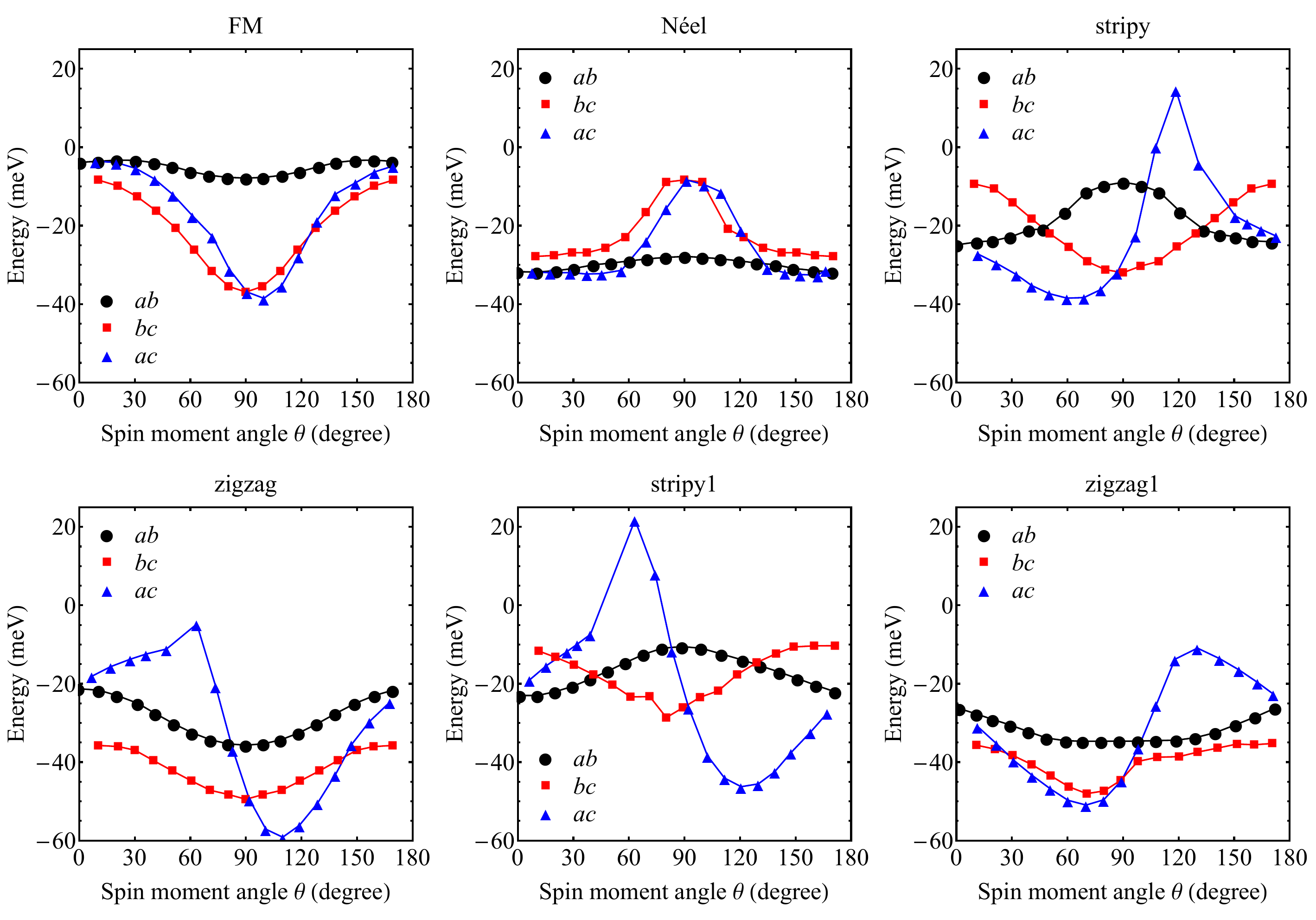}
\caption{The total energies of the indicated magnetic configurations vs. the spin moment direction. The angle $\theta$ is measured from the crystalographic $a$ axis for moments aligned in the $ab$ or the $ac$ planes, and from the crystalographic $b$ axis for moments aligned in the $bc$ plane. A constant energy -139 eV is subtracted from the total energy.}
\label{fig:magnetic}
\end{figure*}

\begin{figure*}
\centering
\includegraphics[width=0.8\textwidth]{band}
\caption{The LDA band structures calculated with different electric dipole configurations. The calculations are performed in a supercell with two layers and four iridium atoms in each layer. The electric dipole configurations are FE$+$FE and FE$-$FE, where the dipoles are ferroelectric within each layer, but the dipoles in adjacent layers are parallel and antiparallel to each other, respectively. Only the 24 energy bands near the Fermi energy are shown, which mainly come from the iridium $t_{2g}$ levels.}
\label{fig:band}
\end{figure*}

\begin{figure*}
\centering
\includegraphics[width=0.8\textwidth]{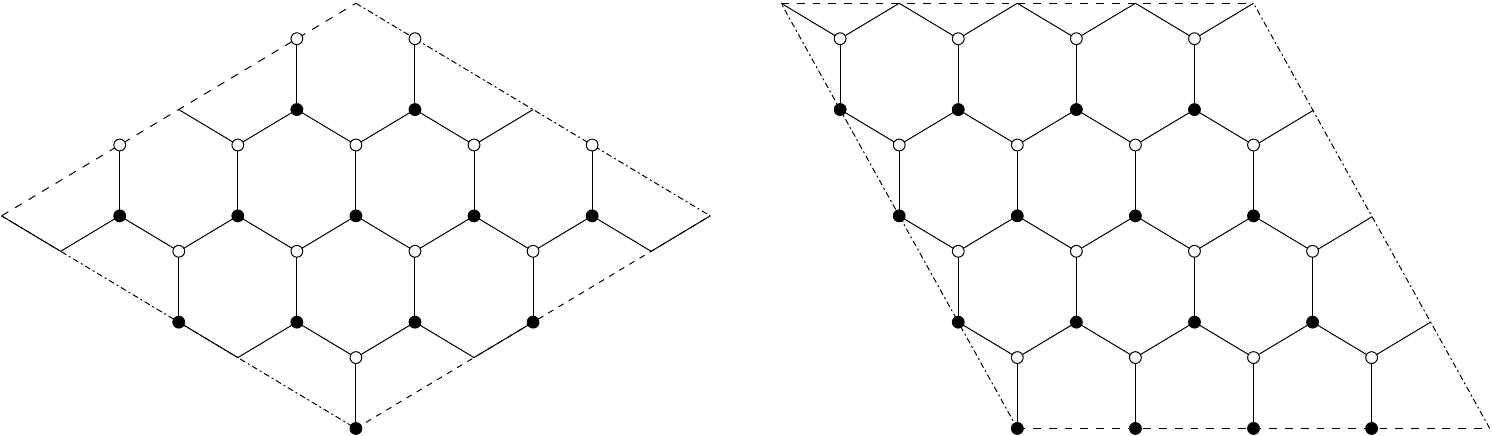}
\caption{(a) The $24$-site lattice. 
(b) The $32$-site lattice.
Closed and open circles indicate two sublattices.}
\label{fig:lattices}
\end{figure*}

\begin{figure*}
\centering
\includegraphics[width=0.96\textwidth]{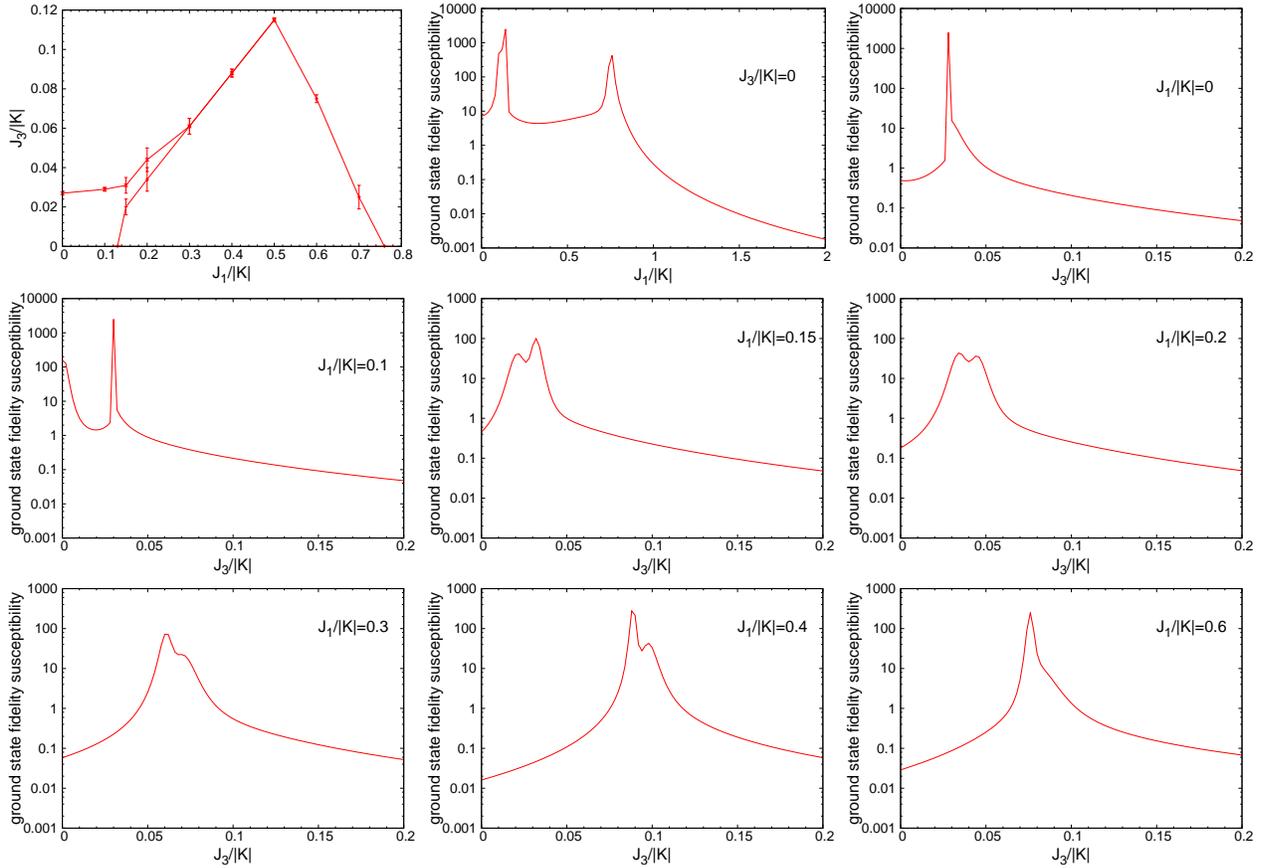}
\caption{(a) The phase boundary determined by exact diagonalization of the $32$-site lattice. The errorbars are estimated from the peak widths of the ground state fidelity susceptibility. 
Lines are guide to the eyes.
(b) Ground state fidelity susceptibility versus $J_1$ for the $32$-site lattice with $J_3=0$.
(c)--(i) Ground state fidelity susceptibility versus $J_3$ for the $32$-site lattice, with $J_1$ values indicated in the figures.}
\label{fig:phase-boundary}
\end{figure*}